%% file: main.tex
\documentclass[acmsmall,screen]{acmart} 
 
\AtBeginDocument{%
  \providecommand\BibTeX{{%
    \normalfont B\kern-0.5em{\scshape i\kern-0.25em b}\kern-0.8em\TeX}}}
    
\setcopyright{rightsretained}
\acmPrice{}
\acmDOI{10.1145/3626465}
\acmYear{2023}
\copyrightyear{2023}
\acmSubmissionID{}
\acmJournal{PACMHCI}
\acmVolume{7}
\acmNumber{ISS}
\acmArticle{}

\usepackage{xspace}
\usepackage{multirow}
\usepackage{calc}

\begin{document}

\input{commands}

\title[BrickStARt: Enabling In-situ Design and Tangible Exploration]{BrickStARt: Enabling In-situ Design and Tangible Exploration for Personal Fabrication using Mixed Reality}

\author{Evgeny Stemasov}
\orcid{0000-0002-3748-6441}
\email{evgeny.stemasov@uni-ulm.de}
\affiliation{%
  \institution{Ulm University}
  \city{Ulm}
  \country{Germany}
}

\author{Jessica Hohn}
\orcid{0009-0007-5548-3803}
\email{jessica.hohn@alumni.uni-ulm.de}
\affiliation{%
  \institution{Ulm University}
  \city{Ulm}
  \country{Germany}
}

\author{Maurice Cordts}
\orcid{0000-0002-1605-2759}
\email{maurice.cordts@uni-ulm.de}
\affiliation{%
  \institution{Ulm University}
  \city{Ulm}
  \country{Germany}
}

\author{Anja Schikorr}
\orcid{0000-0003-0842-5890}
\email{anja.schikorr@alumni.uni-ulm.de}
\affiliation{%
  \institution{Ulm University}
  \city{Ulm}
  \country{Germany}
}

\author{Enrico Rukzio}
\orcid{0000-0002-4213-2226}
\email{enrico.rukzio@uni-ulm.de}
\affiliation{%
  \institution{Ulm University}
  \city{Ulm}
  \country{Germany}
}

\author{Jan Gugenheimer}
\orcid{0000-0002-6466-3845}
\email{jan.gugenheimer@tu-darmstadt.de}
\affiliation{%
  \institution{TU-Darmstadt}
  \city{Darmstadt}
  \country{Germany}
}
\affiliation{%
  \institution{Institut Polytechnique de Paris}
  \city{Paris}
  \country{France}
}

\renewcommand{\shortauthors}{Stemasov, Hohn, Cordts, Schikorr, Rukzio, Gugenheimer}

\begin{abstract}
  \input{sections/abstract}
\end{abstract}



\begin{teaserfigure}
    \centering
    \includegraphics[width=\textwidth]{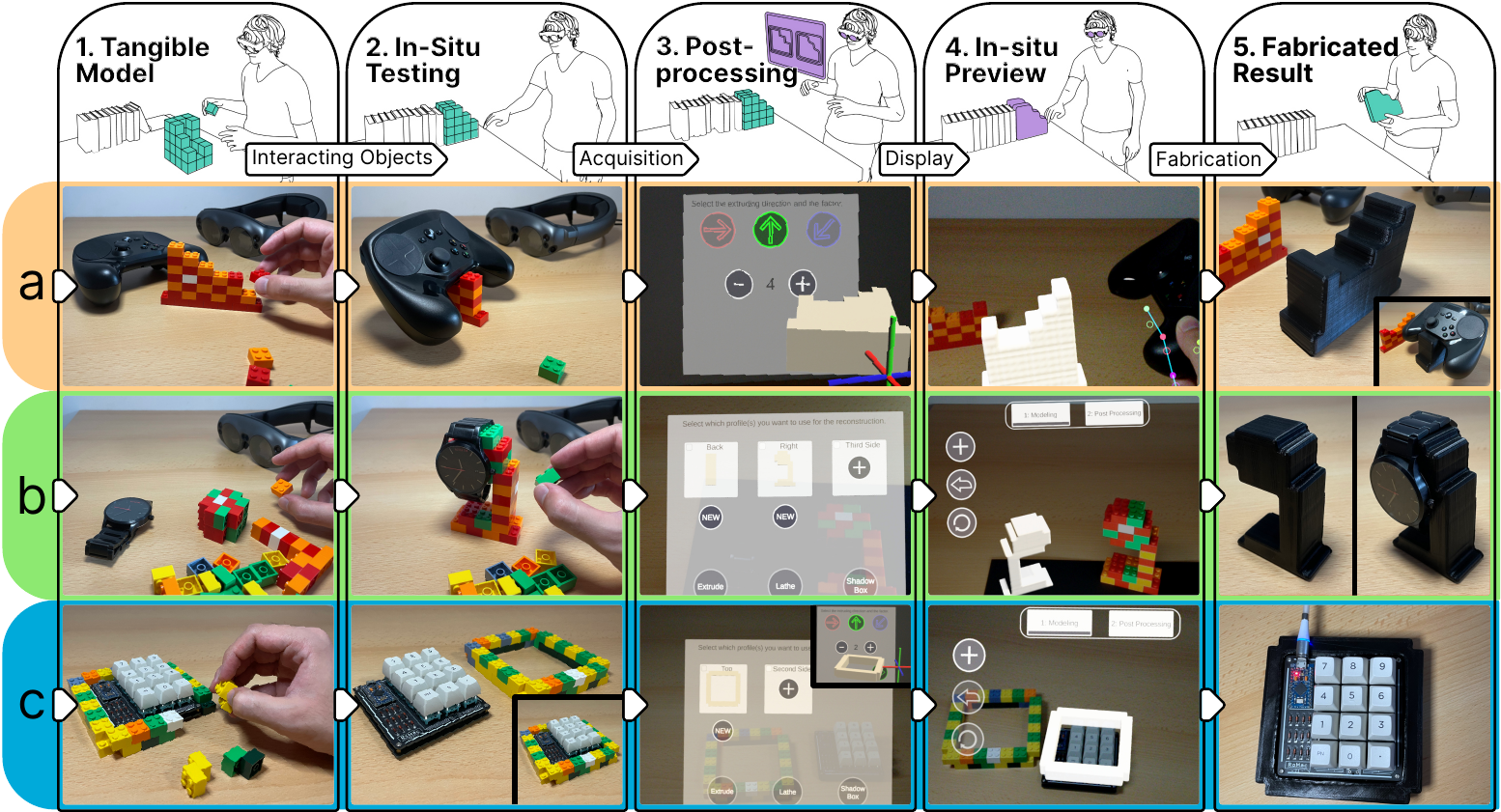}
    \caption{\system embraces in-situ tangible interaction to enable users to design personalized objects, such as stands (a -- controller stand, b -- watch stand) or enclosures (c -- number pad case). Users first create a low-fidelity tangible model using interlocking construction bricks (1). 
    It can be used to test for stability (2a), balance (2b), or fit (2c) early and in-situ. Users can easily and quickly alter this tangible model to fix errors they identified. Then, users scan the model with an AR headset to refine it (3), for instance, to smooth it (3a). The refined result can be previewed in-situ (4) and fabricated (5) to fulfill the user's requirements.}
    \Description[]{Figure 1.: Five components to create a printed 3D model using BrickStARt: Tangible Model, In-Situ Testing, Postprocessing, In-Situ Preview and Fabricated Result. Tangible Model shows a drawn person wearing an AR headset building a brick model on a table. Besides, there are several books, stacked side by side in an unordered fashion. In-Situ Testing shows the same person with the AR headset. Beside the books, the brick model is placed to hold the books. They are now arranged slightly better. Postprocessing shows the same person with the AR headset. A rectangular canvas floating before the person depicts an AR interface. It shows the silhouette of the brick model, one with straight edges, one smoothed. In-situ Preview shows the person wearing the headset, with the smoothed bookend close to the books. Fabricated result shows the printed 3D model of the bookend being held by the person in both hands. Below each of the 5 steps, three examples a, b, and c are shown with photos arranged in rows. Figure 1.a: Image row a shows the process to create a game controller stand. Part 1, Tangible Model, shows a table - on it, a black game controller and the Magic Leap headset in the background. In the foreground, a person is building a model with bricks. Their right hand is visible. Part 2, In-Situ Testing, shows the Magic Leap in the background. In front of it, the controller is balanced on the brick model built in Part 1. Part 3, Postprocessing, shows a screenshot of the BrickStARt application. The brick model was scanned and converted into a digital geometry. Three arrows are shown to extrude the object in an axis: top, right, bottom left. Below there are a plus and a minus to increase or decrease the size.  Part 4, Preview shows a screenshot of the application with the generated geometry augmented over the real environment. In the real environment, there is the controller to compare the size of the geometry with this object. In the background, there is the initial brick model.Part 5, Fabricated Result, shows the printed 3D model of the geometry. Besides, there is the brick model. In the bottom right there is an image of the controller mounted on the printed 3D model. Figure 1.b: Image row b shows the process to create a watch stand. Part 1, Tangible Model, shows a watch and the Magi Leap headset in the background. In the foreground, a person is building two models with bricks. Part 2, In-Situ Testing, shows the Magic Leap in the background. In front of it, the watch is mounted on the brick model built in Part 1. Part 3, Postprocessing, shows a screenshot of the BrickStARt application. The brick models were scanned as the back and right sides of the model. The scans are displayed as geometries labeled Back and Right. Below each geometry, there is a button NEW to rebuild the geometry. With a plus, there is the possibility to add a third side for the model. Below there are the buttons Extrude, Lathe, and Shadow Box. Part 4, Preview, shows a screenshot of the application with the result of Shadow Box of both sides of the scanned models. In the background, the real environment shows the initial brick model to compare it with the generated geometry. Part 5, Fabricated Result shows the printed 3D model of the watch stand, to the left, without the watch present, to the right, with the watch mounted on top. Figure 1.c: Image row c shows the process to create a frame for a Numpad. Part 1, Tangible Model, shows a person building a brick model around a Numpad. Part 2, In-Situ Testing, shows a Numpad and the frame built in Part 1. In the bottom right, the Numpad is inside the brick frame. Part 3, Postprocessing, shows a screenshot of the BrickStARt application. The brick model was scanned as the top view of the model. The scan is displayed as geometry labeled as Top. Below the geometry, there is a button NEW to rebuild the geometry. With a plus, there is the possibility to add a second and third side for the model. Below there are the buttons Extrude, Lathe, and Shadow Box. On the top right image, there are three arrows top, right, and bottom left to extrude the object. A plus and minus below give the possibility to increase or decrease the geometry. The geometry below shows the current size. Part 4, Preview, shows a screenshot of the application with the result of Postprocessing: the extruded casing is overlaid over the numpad. The real environment shows the initial brick model and the Numpad beside. The digital geometry is superimposed over the Numpad to compare the size of the result. Part 5 - Fabricated result shows the numpad inserted in the case. }
    \label{fig:teaser}
\end{teaserfigure}

\maketitle

\input{sections/introduction}
\input{sections/relatedwork}
\input{sections/system}
\input{sections/evaluation}

\input{sections/discussion}

\input{sections/limitations}
\input{sections/futurework}
\input{sections/conclusion}
\input{sections/acknowledgements}

\bibliographystyle{ACM-Reference-Format}
\bibliography{zotero}

\end{document}

%% file: commands.tex
\def\UrlBreaks{\do\/\do-\do\_}

\newcommand{\todo}[1]{}
\renewcommand{\todo}[1]{{\color{red} \textbf{TODO: {#1}}}}

\newcommand{\fix}[1]{}
\renewcommand{\fix}[1]{{\color{orange} \textbf{FIXME: {#1}}}}

\newcommand{\cit}[1]{}
\renewcommand{\cit}[1]{{\color{green} \textbf{CITEME: {#1}}}}

\newcommand{\system}{BrickStARt\xspace}
\newcommand{\tk}{Tinkercad\xspace}
\newcommand{\systemEmph}{\emph{\system}\xspace}
\newcommand{\systemEmphSp}{\systemEmph\xspace}

%% file: sections/abstract.tex
3D-printers enable end-users to design and fabricate unique physical artifacts but maintain an increased entry barrier and friction.
End users must design tangible artifacts through intangible media away from the main problem space (ex-situ) and transfer spatial requirements to an abstract software environment.
To allow users to evaluate dimensions, balance, or fit early and in-situ, we developed \system, a design tool using tangible construction blocks paired with a mixed-reality headset. 
Users assemble a physical block model at the envisioned location of the fabricated artifact.
Designs can be tested tangibly, refined, and digitally post-processed, remaining \emph{continuously in-situ}.
We implemented \system using a Magic Leap headset and present walkthroughs, highlighting novel interactions for 3D-design.
In a user study ($n=16$), first-time 3D-modelers succeeded more often using \system than Tinkercad.
Our results suggest that \system provides an accessible and explorative process while facilitating quick, tangible design iterations that allow users to detect physics-related issues (e.g., clearance) early on. 

%% file: sections/introduction.tex
\section{Introduction}\label{sec:introduction}
    Digital Fabrication can provide unrivaled, industry-grade precision to end-users and consumers, allowing them to create highly personalized physical objects.
    This is described in the vision of personal fabrication~\cite{gershenfeldDesigningRealityHow2017,gershenfeldFabComingRevolution2005,baudischPersonalFabrication2017}. 
    Proficient users may then design and fabricate decorative artifacts tailored to their taste (e.g., vases or ornaments), functional artifacts tailored to their mechanical requirements (e.g., mounts, attachments, spare parts), or any combination of those.
    In the future, given sufficiently powerful fabrication processes, \emph{any} requirement for a physical artifact may be solvable quickly and satisfyingly through personal (digital) fabrication.
    
    While personal digital fabrication makes these compelling promises (e.g., precision, personalization), it remains a tool used mainly by hobbyists (who were involved in -- analog/digital -- craft anyway) or technology enthusiasts (intrinsically motivated to engage with the technology)~\cite{stemasovRoadUbiquitousPersonal2021,hudsonUnderstandingNewcomers3D2016}.
    This can be ascribed to various types of \emph{friction} users have to overcome during design and fabrication: most design tools have to be learned, and involve complex, lengthy processes of design and iteration.
    In particular, design tools are used \emph{ex-situ}, away from the problem space~\cite{ashbrookAugmentedFabricationCombining2016}, are \emph{intangible} (despite being used for creating \emph{tangible} artifacts), and the exploration of physical consequences is left to either complex simulation software or fabricated iterations of the design (i.e., engaging in physical prototyping). 
    We see a set of specific aspects to address here: 1) the design of (future / ultimately) tangible objects is mostly intangible; 2) the design of these future objects happens ex-situ, instead of in-situ; 3) users are restricted to exploring physical consequences either with a finished design (to simulate \emph{digitally}) or after fabrication (to evaluate \emph{physically}).
    
    We propose \system\footnote{Naming derived from Brick + Kickstart + AR}, a mixed reality 3D-modeling tool supported by tangible elements (fig. \ref{fig:teaser}), implemented for the Magic Leap One\footnote{\url{https://ml1-developer.magicleap.com/en-us/home}, Accessed: 8.5.23} head-mounted display (HMD).
    With \system, users start their design by coarsely approximating their desired shape and features through colored, interlocking plastic bricks (e.g., LEGO\textregistered, fig. \ref{fig:teaser}.1).
    Users may either build their entire design (fig. \ref{fig:teaser}b) or define separate outlines (1 brick ``deep'') to alter later on (fig. \ref{fig:teaser}a).
    The brick model enables them to coarsely explore physical, tangible consequences of their design, such as the stability of their object (fig. \ref{fig:teaser}.1b) early in their design process.
    This also provides a worst-case estimate of fit (fig. \ref{fig:teaser}.2c), balance (fig. \ref{fig:teaser}.2b), or clearance.
    \system offers a set of methods to alter and refine the design in mixed reality, like extrusion (fig. \ref{fig:teaser}.3a), smoothing, and basic geometrical operations. 
    Afterward, the design can be previewed in-situ (fig. \ref{fig:teaser}.4) and altered further (e.g., smoothed).
    Lastly, users may fabricate the artifact they designed and use it within the physical context they explored and made their design in (fig. \ref{fig:teaser}.5).
    With \system, we focus on designing everyday household artifacts~\cite{shewbridgeEverydayMakingIdentifying2014}, over complex engineering challenges, as these household artifacts are an aspect of widespread adoption of personal fabrication and its embedding and increasing relevance in everyday life~\cite{stemasovRoadUbiquitousPersonal2021}.
    These household artifacts, such as jigs, mounts, or decorative vases, may not demand the highest precision, but benefit from users expressing their uniquely personal requirements, both functional and aesthetic, and especially in relation to their unique personal contexts.
    Established CAD (computer-aided-design) software, along with prevalent paradigms of 3D-modeling~\cite{stemasovRoadUbiquitousPersonal2021}, do not necessarily fit this use-case, as they often focus on offering users the highest possible precision.
    
    We explored the use of \system in a user study ($n=16$).
    We found that users immediately got started designing the tangible model and were able to evaluate how a design would behave early on.
    This happened through tangible, hands-on explorations with the bricks and interacting objects, but also visually, using the digital model displayed in-situ through the HMD.
    Users with no prior 3D-modeling experience created fewer designs that did not comply with balance or clearance requirements using \system, compared to Autodesk \tk, indicating the suitability of the paradigm for novice 3D modelers.
    While \system is not meant to provide users with unlimited precision and fidelity, it offers them a tangible, in-situ design process that is particularly applicable to everyday household items that may interact with existing counterparts in a physical environment~\cite{ashbrookAugmentedFabricationCombining2016,mahapatraBarriersEndUserDesigners2019}.
    Such household items, like planters or stands for various objects, are well-represented in the popular designs in repositories like Thingiverse\footnote{\url{https://www.thingiverse.com}, Accessed 2.5.23}.
    Unlike the use of Thingiverse or desktop-based CAD, \system situates all \emph{relevant} steps (i.e., design and exploration) of a personal fabrication process within the physical context of the future artifact and in conjunction with existing interacting objects, allowing users to test and iterate while designing (fig. \ref{fig:teaser}.2).
    This enables more dynamic explorations and iterations of a design, without having to either 1) employ a fabrication device to \emph{permanently} manufacture a prototype; or 2) transfer all information from the physical environment to set up a digital design and simulation, risking errors along the way.
    We argue that this is necessary to enable novices, inexperienced end-users~\cite{ashbrookAugmentedFabricationCombining2016,mahapatraBarriersEndUserDesigners2019}, or even non-users~\cite{stemasovRoadUbiquitousPersonal2021} to start exploring and engaging with personal fabrication in the future.
    
    The contributions of this work are:
    \begin{itemize}
        \item The\textbf{ concept of using tangible elements augmented with a mixed reality headset to define artifacts for personal fabrication}. This combination enables physical, hands-on exploration prior to fabrication while allowing users to continuously interact in-situ (i.e., in the object's future context). We further present the prototype implementation of \system, a mixed reality modeling tool embracing this approach.
        \item \textbf{Insights gathered from a user study of \system ($n = 16$) and application walkthroughs}, not only confirming the feasibility of our approach but also showing that \system may prevent novice users from making errors referring to measuring (clearance) and creating stable designs (balance) by allowing hands-on explorations without having to fabricate prototypes. 
    \end{itemize}

%% file: sections/relatedwork.tex
\section{Related Work}\label{sec:rw}
    \system is inspired by and related to 3 core directions: tangible computer-aided design (CAD) tools, 2D-to-3D CAD workflows, and in-situ design tools.
    Our focus lies on the \emph{design} of artifacts and less on their fabrication, even if the brick assembly could be considered a fabrication method~\cite{muellerFaBrickationFast3D2014}.
    With \emph{in-situ modeling or interaction}, we focus on the location of \textbf{future use} matching the location of design, not the location of fabrication.
    \system also aims to be an in-situ design tool that makes the context (i.e., interacting objects), but also the workpiece tangible.
    Ideally, a tool provides both tangibility and situatedness, to allow for quick in-situ design explorations~\cite{agrawalProtopiperPhysicallySketching2015}, early testing, and easier transfer of requirements (e.g., sizes) between physical context and digital model~\cite{mahapatraBarriersEndUserDesigners2019}.

    \subsection{Tangible CAD}
        Adding tangibility to computer-aided design tools introduces a perceptual dimension lost with most screen-based CAD systems.
        This enhances users' understanding of the physical properties a (future) artifact may exhibit.
        Prior works have explored tangible systems for both \emph{design} and \emph{fabrication}.
        Savage et al. allowed users to model in clay or other materials~\cite{savageMakersMarksPhysical2015}.
        Stickers were used to annotate where the system should insert mounts for functional elements like buttons~\cite{savageMakersMarksPhysical2015}.
        Jones et al. presented a comparable approach for interactive device design based on clay combined with placeholders for interactive elements~\cite{jonesWhatYouSculpt2016}.
        CopyCAD by Follmer et al. enabled users to take real-world objects to scan and embed in digital designs~\cite{follmerCopyCADRemixingPhysical2010}, which was further explored with children as a user group in KidCAD~\cite{follmerKidCADDigitallyRemixing2012}.
        Various approaches have previously focused on a ``building-block''-like interaction for 3D design. 
        This constrains the users' input to a fixed grid, in contrast to free-form approaches based on clay.
        StrutModeling by Leen et al. is a construction kit out of struts and hubs, enabling users to prototype objects in situ and in a tangible fashion~\cite{leenStrutModelingLowFidelityConstruction2017}.
        However, their approach moves steps like postprocessing and previewing of post-processed models ex-situ (i.e., to a desktop computer).
        Anderson et al. presented a 3D-modeling approach based on tangible blocks or clay, which are scanned and interpreted to be rendered as a digital model~\cite{andersonTangibleInteractionGraphical2000}.
        Their work is predated by comparable approaches focused on architectural design~\cite{aishArchitectureNumbersCAAD1984}.
        Dynablock by Suzuki et al. presented ``dynamic 3D printing'', an approach to fabricate reconfigurable, low-fidelity 3D models~\cite{suzukiDynablockDynamic3D2018}, which allows for an overlap of design and fabrication.
        FaBrickation by Mueller et al. used building blocks to accelerate \emph{fabrication} and therefore prototyping cycles~\cite{muellerFaBrickationFast3D2014}, covering a different phase of the personal fabrication process than \system, which is focused on accessible design over quick prototype manufacturing.
        In the context of electronics design, ElectronicsAR supported novices  designing circuits~\cite{fegerElectronicsARDesignEvaluation2022}, and NatCut facilitated the design of enclosures~\cite{schneegassNatCutInteractiveTangible2014} through tangible interaction and augmented information.
        
        With \system, we similarly take a tangible approach to design for fabrication but want to situate all relevant steps of a design in the design's relevant physical context (i.e., where it will be \emph{used}), unifying digital and physical spaces~\cite{doganFabricateItRender2022}.
        We furthermore embrace a lower fidelity and resolution, while still focusing on the objects' interactions with the physical environment.
        This tangible exploration with a low-fidelity model allows for early tests of physical properties.
        Meanwhile, actions involving leaving the block grid, altering the model further, and creating smooth surfaces are still enabled and explorable in-situ, in mixed reality.

    \subsection{2D-to-3D CAD Workflows}
        Both academic works and established CAD software have allowed users to define features in 2D to convert them to 3D features later (e.g., a 2D sketch that is extruded to create a 3D volume).
        To simplify the design process, \system also allows for this kind of workflow, in addition to the assembly of complete 3D objects.
        An influential example is Teddy by Igarashi et al., which inflates 2D outlines -- drawn by novice users -- into 3D models~\cite{igarashiTeddySketchingInterface1999}.
        SketchChair by Saul et al. took a similar approach within the context of chair design~\cite{saulSketchChairAllinoneChair2011}. 
        Users began by sketching the desired outline and verifying the design through simulation afterward. 
        From this design, a cutting plan for a 3D chair is generated.
        McCrae et al. presented FlatFitFab, which focuses on combining planar sections to 3D objects ~\cite{mccraeFlatFitFabInteractiveModeling2014}.
        The tool further supports users by providing physics simulations, for instance, by accounting for loads put onto an assembly of planar sections.
        CutCAD is an open-source tool enabling users to create 3D objects based on assemblies of 2D parts~\cite{hellerCutCADOpensourceTool2018}.
        
        Tools that focus on fabrication or overlap between fabrication and design have also been present in the literature. 
        LaserStacker~\cite{umapathiLaserStackerFabricating3D2015} and LaserOrigami~\cite{muellerLaserOrigamiLasercutting3D2013} are both tools to create 3D geometry with 2D cutting machines, leveraging their outstanding speed in planar cutting.
        StackMold similarly focuses on fabrication, but uses stacked layers of 2D molds to create 3D objects~\cite{valkeneersStackMoldRapidPrototyping2019}.
        HingeCore~\cite{abdullahHingeCoreLaserCutFoamcore2022} and FoolProofJoint~\cite{parkFoolProofJointReducingAssembly2022} similarly focus on the assembly of cut 2D elements to 3D objects.
        
        In a similar fashion to these works, we want to benefit from 2D features for the design of 3D objects.
        We rely on metaphors established in CAD software, which allows users to \emph{partially} define and explore objects and focus on relevant features.
        We furthermore want users to benefit from tangibility and an in-situ process while focusing on design instead of fabrication.
        Unlike FlatFitFab~\cite{mccraeFlatFitFabInteractiveModeling2014} or CutCAD~\cite{hellerCutCADOpensourceTool2018}, we aim to allow users to \emph{explore} their design and its physical properties in a lower resolution, and tangible fashion which is quick to iterate and change without demanding material use.
    
    \subsection{In-situ Design and Fabrication Tools}
        We lastly consider \system to be an in-situ design tool, which situates design and iteration within an artifact's future context.
        Comparable tools are particularly relevant in the context of design for fabrication. 
        This allows users to tangibly experience both the workpiece and the context it will interact with after fabrication.
        Prior examples of in-situ \emph{design} tools, for instance, include MixFab by Weichel et al.~\cite{weichelMixFabMixedrealityEnvironment2014}, which is a self-contained design environment using mixed reality. 
        Users can introduce real-world objects into the system to alter geometry with their help~\cite{weichelMixFabMixedrealityEnvironment2014}.
        Lau et al. presented Modeling in Context, which is a design tool leveraging tangible primitives for in-situ 3D-modeling~\cite{lauModelingincontextUserDesign2010}.
        Tangible Version Control by Letter and Wolf explored versioning of physical artifacts supported with an HMD~\cite{letterTangibleVersionControl2022}.
        DesignAR by Reipschläger and Dachselt~\cite{reipschlagerDesignARImmersive3DModeling2019} combined an HMD (head-mounted display) with an interactive surface, which for instance, allows users to trace reference objects and features.
        AutomataStage by Jeong et al. focused on creativity support in the context of designing automata~\cite{jeongAutomataStageARmediatedCreativity2023}.
        
        In the context of (personal) fabrication, which focuses on the creation of tangible artifacts, in-situ \emph{fabrication} tools are also present in literature, often combining the location of design with the location of fabrication.
        WireDraw presented an approach with instructions shown through an HMD guiding a user's manual fabrication process with a 3D-printing pen~\cite{yueWireDraw3DWire2017}.
        RoMA by Peng et al. couples design and fabrication, allowing manual input and fabrication by the user, with the option to introduce existing objects into the device's space~\cite{pengRoMAInteractiveFabrication2018}.
        Protopiper by Agrawal et al. can be considered a low-fidelity fabrication tool at a larger scale than most other approaches~\cite{agrawalProtopiperPhysicallySketching2015}, and likened to tools for architecture-scale design support~\cite{mitterbergerAugmentedBricklaying2020,mitterbergerExtendedRealityCollaboration2023,mitterbergerAugmentedHumanExtended2022}. 
    
        As an alternative to processes of design or fabrication, remixing~\cite{roumenGrafterRemixing3DPrinted2018,stemasovImmersiveSamplingExploring2023} and retrieval~\cite{stemasovMixMatchOmitting2020,stemasovShapeFindARExploringInSitu2022} of existing (e.g., crowdsourced) designs have also been considered in prior works. 
        CustomizAR by Liang et al. further expanded upon this notion by supporting users in discovering parametric designs and measuring real-world objects to customize the models~\cite{liangCustomizARFacilitatingInteractive2022}. 
        In comparable fashion to greatly simplified modeling tools, they lower the effort required to learn and benefit from (3D-) modeling tools.

        With \system, we embrace these notions but want to make both the user's model and interacting workpieces (i.e., the physical context) simultaneously tangible.
        This enables early testing and explorations, ideally reducing design errors that demand iterations.
        We furthermore want users to continuously interact in-situ with the object's future physical context, and not a decoupled standalone fabrication device or design workstation.

%% file: sections/system.tex
\section{BrickStARt}\label{sec:system}
    The following sections outline the design rationale for \system, our envisioned interaction flow for the tool, and describe the application itself.
    Based on the literature outlined in \autoref{sec:rw}, we aimed for the following design goals: 1) \textbf{tangible interaction for modeling}~\cite{horneckerGettingGripTangible2006,andersonTangibleInteractionGraphical2000} that enables early checks of physics (in a quick, intuitive fashion~\cite{jacobRealitybasedInteractionFramework2008}); 2) \textbf{fast explorations/iterations} before any digital design is attempted and before any permanent prototype is manufactured; 3) \textbf{continuous in-situ interaction} that does not require leaving a physical context while designing for it~\cite{stemasovMixMatchOmitting2020}.
    These aspects are meant to enable tangible, in-situ modeling, while simultaneously lowering the skill floor for 3D-design of physical objects (i.e., easy entry for novices~\cite{yehCraftML3DModeling2018}). 
    We consider straightforward problems and similarly straightforward geometries to be particularly relevant for novices~\cite{alcockBarriersUsingCustomizing2016,stemasovRoadUbiquitousPersonal2021,baudischKyub3DEditor2019,liuAngleCADSurfacebased3D2022}, and want to emphasize that using desktop-based, ``traditional'' CAD tools for such tasks can prove to be a non-trivial task for genuine novices to the space~\cite{mahapatraBarriersEndUserDesigners2019}.
    Focusing on relatively simple geometries may not leverage the full potential of digital fabrication, but still provides meaningful benefits in terms of creative expression and the design and acquisition of physical artifacts that are uniquely tailored to an end user's requirements and context~\cite{raynaRapidPrototypingHome2016,stemasovRoadUbiquitousPersonal2021}.
    With \system, we support this by employing low-fidelity tangible modeling, which enables early exploration of physical consequences, ideation, and iteration.
    This low-fidelity tangible model can then be refined and altered through a digital system, while still remaining in-situ (i.e., at the location of the artifact's future \emph{use}), reducing the potential for errors and effort in transfers between digital design and physical context~\cite{mahapatraBarriersEndUserDesigners2019}.

    \subsection{Design Rationale and Interaction Flow}\label{sec:rationale}
        The following paragraphs elaborate on the design decisions we made for the development of \system, focusing on material choice, the hardware setup, and the intended context of use.
            
        \paragraph{Mixed Reality for Design}
            For \system, we further rely on the means of spatial computing (i.e., a mixed reality head-mounted display -- HMD).
            This allows \system to operate \textbf{continuously in-situ}: both in the tangible design stage, but also in the refinement stage later on. 
            At no point in the design process, do users have to engage in transfers of requirements (cf.,~\cite{ashbrookAugmentedFabricationCombining2016,mahapatraBarriersEndUserDesigners2019} to a (spatially) separate workspace location, but rather interact in-situ.
            This is meant to reduce friction, by either omitting or simplifying the transfers between physical (problem) space and (digital) design space.
            This includes measures (i.e., transferring an object's size to the design tool~\cite{mahapatraBarriersEndUserDesigners2019}), but also previewing (i.e., transferring a digital design to the problem space~\cite{stemasovMixMatchOmitting2020}). 
            While both directions do not exhibit perfect precision, they can help detect or avoid errors early on, instead of doing so after fabrication, thereby saving material.
            We further consider spatial computing to be an emerging sensor platform, able to provide a variety of inputs to the design process (e.g., depth scanning, context understanding) in the future.
            Using a headset, specifically, further supports exploration/iteration: users can start working on a new brick design, while still seeing a previous, digital version of it -- both situated in the same physical context.
            We consider the use of mixed/augmented reality the crucial part in this case, and less the specific hardware platform (e.g., HMD vs. phone), as long as they fulfill the outlined designed goals. 
            Using a tablet or a phone to provide AR functionality would work in a similar fashion (e.g., as shown in~\cite{liangCustomizARFacilitatingInteractive2022}) and may yield different tradeoffs (e.g., occupation of hands, access to required hardware).
            HMDs also provide stereoscopic depth cues, which are particularly influential at close range~\cite{cuttingPerceivingLayoutKnowing1995} and relevant when the device has to be hand-held while interacting~\cite{gombac3DVirtualTracing2016}.
            We further combine gesture-based and controller-based inputs in our implementation of \system (cf.,~\cite{huangEvaluationHybridHand2021}), to enable a separation of concerns between UI interactions and interactions made with 3D models in the application. 
            The fundamental benefits of mixed/augmented reality are conceptually similar across these platforms and would stay the same, even with increasing miniaturization.
            
        \paragraph{Usage Context}
            With \system, we focus on a \textbf{subset of users} to support and a \textbf{subset of objects} to be designed: we focus on current non-users of digital/personal fabrication~\cite{stemasovRoadUbiquitousPersonal2021}, or novice users~\cite{hudsonUnderstandingNewcomers3D2016}.
            Their requirements in the context of personal fabrication may differ from the ones put forward by expert users~\cite{yildirimDigitalFabricationTools2020,gulayIntegratedWorkflowsGenerating2019}, and especially from requirements of enthusiast users~\cite{hudsonUnderstandingNewcomers3D2016,stemasovRoadUbiquitousPersonal2021}.
            For these users, searching for and customizing objects found on platforms like Thingiverse is a valid alternative, but not always possible or trivial to do~\cite{alcockBarriersUsingCustomizing2016,stemasovRoadUbiquitousPersonal2021}, as their intended object may not be available or customizable~\cite{alcockBarriersUsingCustomizing2016}.
            Similarly, customization of a parametric design to the user's requirements can similarly prove to be non-trivial, as users still have to \emph{correctly} acquire and transfer requirements from the physical context to the design~\cite{mahapatraBarriersEndUserDesigners2019,liangCustomizARFacilitatingInteractive2022,ashbrookAugmentedFabricationCombining2016,kimUnderstandingUncertaintyMeasurement2017}.
            In turn, this user group may engage with personal fabrication in a different way: focused on outcomes, and less on the process~\cite{hudsonUnderstandingNewcomers3D2016}, and focused on small, everyday household artifacts~\cite{shewbridgeEverydayMakingIdentifying2014,wadeChallenges3DPrinting2017} instead of complex engineering challenges~\cite{stemasovRoadUbiquitousPersonal2021}.
            Geometry-wise, this implies that we focus on artifacts that, for the current version of \system, are small-scale,  do not require features below 8mm\footnote{approximately half a brick wide} in size, and are solutions to either mechanical/physical problems or fulfill aesthetic functions in users' households.
            We elaborate on the limitations of the approach in section \ref{sec:limitations}, but emphasize that \system is powerful enough for everyday household artifacts, as indicated by the study and the use-cases presented across sections \ref{sec:study} and \ref{sec:walkthroughs}.

        \paragraph{Material}
            We settled on the use of interlocking plastic bricks (e.g., LEGO\textregistered), after considering alternatives like paper (to sketch on or to fold), malleable plastics (e.g., as used in~\cite{taniguchiRapidPrototypingSystem2018}) or clay.
            Alternatives to these plastic bricks, such as modeling clay~\cite{savageMakersMarksPhysical2015,weichelReFormIntegratingPhysical2015}, are present in the literature.
            However, interlocking plastic bricks provide a set of advantages, which we deemed appropriate for \system: 
            1) The \textbf{learning effort} that is required to use them successfully, even to solve mechanical problems, is minimal -- they can almost be considered a universal language for expression, that is even suitable for children~\cite{rubensFlyingLEGOBricks2020}, and their use has been proposed for learning CAD~\cite{yip-hoiTeachingCADModeling2011} as well.
            2) The material properties are comparable to fabricated (i.e., 3D-printed) objects in terms of density and, by extension, physical behavior. 
            While this similarity does not constitute fabrication-awareness~\cite{weichelSPATASpatioTangibleTools2015,wesselyShapeAwareMaterialInteractive2018}, it allows users to tangibly experience and understand how the fabricated result may physically behave and provide a worst-case estimate of this behavior.
            3) The bricks are particularly \textbf{easy to reconfigure} and iterate with while maintaining appropriate structural stability.
            They are also inherently reusable without meaningful degradation over time.

        \paragraph{Process}
            \begin{figure}[ht]
                \centering
                \includegraphics[width=\linewidth]{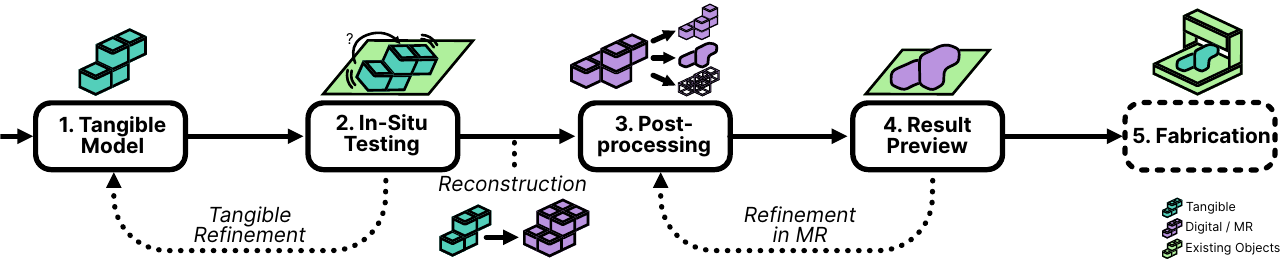}
                \caption{Envisioned process for \system. Users start with a requirement and express it through an initial tangible model (1), which can be tested for physical aspects such as stability (2). This model is then reconstructed for further processing in AR (3). Altered results can similarly be previewed in-situ (4) and fabricated (5), if the result is deemed appropriate. With \system, we aim to avoid iterations with fabricated artifacts (e.g., $5 \rightarrow 2$), but rather iterate and explore early and in-situ ($2 \rightarrow 1$ or $4 \rightarrow 3$).}
                \Description[]{Figure 2.: A flowchart-like diagram with five components: Tangible Model, In-Situ Testing, Postprocessing, Result Preview, and Fabrication. Tangible Model shows a modelled S-block. This feeds into In-Situ Testing illustrated by the S-block as moving. This feeds again in Tangible Model as Tangible Refinement and in Postprocessing with the intermediate step Reconstruction which is illustrated with the S-block which is widened. Postprocessing shows a block model which is smoothed on the left and on the right side the object with lattice reduce. This leads into Result Preview which shows the smoothed Object on a green ground. It feeds into Postprocessing with Refinement in MR and in Fabrication which is illustrated by an 3D-Printer with the object. }
                \label{fig:process}
            \end{figure}
        
        The envisioned process for \system is outlined in figure \ref{fig:process}.
        Our focus lies on the \emph{design} of physical artifacts for personal fabrication, as this is the step where a user's personal input (i.e., their physical context and their requirements) is highly relevant.
        A user with access to personal fabrication technology (either through hardware at home, in public spaces like makerspaces~\cite{garcia-ruizFabLabsRoadDistributed2022,garnierMakingHackingCoding2021}, or through a print service~\cite{bermanAnyoneCanPrint2020}) starts with a certain requirement or desire for a new, physical artifact. 
        This requirement is formulated and evaluated within its own physical context (i.e., where it will reside after fabrication). 
        The user then constructs a tangible model out of, in our case, interlocking building blocks (e.g., LEGO\textregistered).
        This tangible model is meant to be a lower-fidelity approximation of the desired artifact figure (fig. \ref{fig:process}-1), or 2D profiles representing relevant geometric features.
        With this tangible model, the user can already evaluate certain aspects of their design: proportions with respect to the physical context, fit/clearance, or balance/stability (fig. \ref{fig:process}-2).
        If the user notices flaws, the tangible model can be quickly altered, as afforded by the building blocks.
        After sufficient refinement, the user can now use \system to reconstruct the tangible model digitally.
        The user photographs the profile or object with the help of the HMD.
        From this photograph, 3D geometry is inferred and reconstructed.
        \system offers a set of methods to generate (3D) geometry based on a segmentation of 0-3 images.
        This geometry can then be postprocessed, for instance, to generate a smoothed surface (fig. \ref{fig:process}-3).
        The resulting model can be further altered (e.g., by adding primitives like cubes) and previewed in-situ (fig. \ref{fig:process}-4).
        If the user considers the result to be promising, the model can be handed off for fabrication.
        We do not assume any \emph{specific} devices or services for fabrication to be particularly relevant for \system, but chose 3D printing as a suitable and currently available means for fabrication.
        Ultimately, a wide range of users may care more about results and less about the technological means enabling them~\cite{stemasovRoadUbiquitousPersonal2021,hudsonUnderstandingNewcomers3D2016}.
    
    \subsection{\system Application}
        \system was developed using Unity 2019.3 and Lumin SDK 0.24.1 for the Magic Leap One headset.
        Furthermore, the geometry3Sharp\footnote{\url{https://github.com/gradientspace/geometry3Sharp}, Accessed: 5.5.23} and OpenCV for Unity\footnote{\url{https://enoxsoftware.com/opencvforunity/}, Accessed: 4.5.23} packages were used for geometry and image processing respectively.
        For the tangible building blocks, we initially settled on standard 2x2 interlocking plastic toy bricks.
        They are almost cube-shaped, with a size of approximately 15.8 x 15.8x x 11.4 mm, including the studs.
        Later, 4x2 and 8x2 were added, to reduce the influence of having a single block size on users' designs and approaches.
        The tangible building blocks cover the first phases of our envisioned process (fig. \ref{fig:process}-1 \& \ref{fig:process}-2), while the AR application covers the later phases of the process (fig. \ref{fig:process}-3 \& \ref{fig:process}-4).
        The transfer from the tangible model to a digitized one happens through an image processing pipeline, which we describe next.

        \subsubsection{Image Processing}
            \begin{figure}[ht]
                \centering \includegraphics[width=\linewidth]{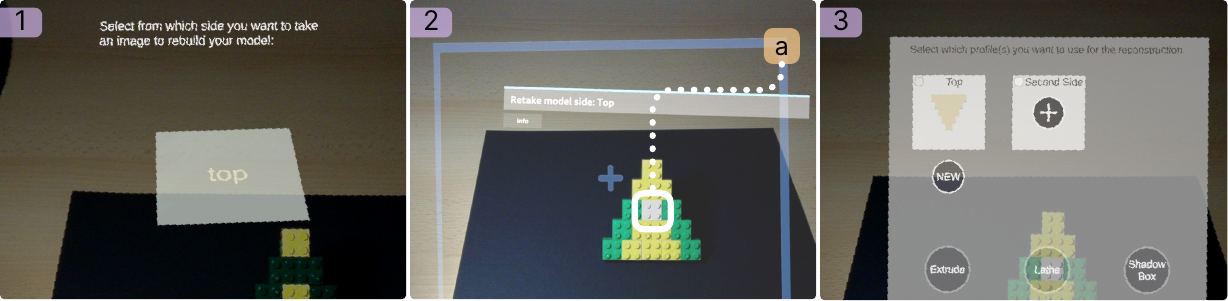}
                \caption{To digitize the tangible model, the user frames it with the headset's camera and takes a picture (1). \system requires users to present the model on a black background and embed a single white brick in their model (2-a) to calibrate the grid size. The result is presented to the user (2). If the user confirms the photograph, the geometry is extracted and presented in a UI where further side-views can be captured (3), or re-attempted}
                \Description[]{Figure 3.: Image 1) shows a black background and a brick model of the real environment. The AR environment shows a cube with selected top side and the instruction “Select from which side you want to take an image to rebuild your model”. Figure 3.1: Image 1) shows a black background and a brick model of the real environment. The AR environment shows a cube with selected top side and the instruction “Select from which side you want to take an image to rebuild your model”. Figure 3.2: Image 2) shows a black background of the real environment. The AR environment shows a photo of the Brick model with reference line to the white brick. Figure 3.3: Image 3) shows a black background and a brick model of the real environment. The AR environment shows on the top left the generated geometry of the model is shown. Right of it a plus is presented to add a new side. At the bottom there are three buttons Extrude, Lathe, and Shadow Box. }
                \label{fig:photo-ui}
            \end{figure}
            
            We chose an image-based approach to digitize the tangible model.
            Current AR/VR headsets possess depth cameras for tracking.
            However, the ones in contemporary mixed reality devices generally deliver a mesh too coarse for the scale \system is working at.
            Users initially place their model on a black background and choose a ``side'' they are acquiring the object from (fig. \ref{fig:photo-ui}-1).
            Then, the headset guides them to frame and photograph their model (fig. \ref{fig:photo-ui}-2).
            The resulting picture is then shown to the user.
            If the user confirms that they have captured an appropriate view of their model (e.g., the model is within the frame and is not occluded or distorted), the processing pipeline is started.
            The image is cropped to match the aspect ratio of the device's field of view.
            It is then scaled down to a resolution of 300x255 pixels, to accelerate processing steps later in the pipeline.
            For the following steps, OpenCV for Unity 2.3.9 was used, which in turn uses OpenCV 4.3.0 internally.
            The scaled-down image is denoised and a Gaussian blur is applied.
            Afterward, mean shift filtering is used for image segmentation, which quantizes the number of colors\footnote{As the total number is unknown (i.e., users are free to use any number of colors greater than 2 in their tangible model), k-means is not applicable here.}.
            To extract salient edges from the image, canny edge detection is used.
            The retrieved edges serve as a region of interest for the GrabCut algorithm~\cite{rotherGrabCutInteractiveForeground2004}, which separates the image into a foreground and background.
            To infer a scale of the tangible model a user has built, \system requires them to use a single white brick in their construction (fig. \ref{fig:photo-ui}-2a).
            The brick has to be visible once per photo (e.g., one brick visible from each perspective or multiple bricks, each visible from a single side only).
            This allows all other bricks to have any other color.
            Through this single brick, the x and y sizes of the brick grid are inferred.
            This reference brick could also be simply placed within the photographed frame or be added to the model and removed by the system later on.
            However, we found that using the reference brick as an actual part of the model avoided distortions introduced through the camera.
            After the size of the grid has been determined, \system generates a binary bitmask from the image, based on possible (i.e., valid) locations of bricks within the captured frame.
            Based on this bitmask, a digital copy of the scanned tangible model's outline is created  (fig. \ref{fig:photo-ui}-3).
            This enables users to apply specific reconstruction methods to create 3D models from the outlines. 
        
        \subsubsection{Reconstruction}
            Having acquired one or more profiles (or sides of a model), users can now reconstruct it to a digital model to alter, postprocess, and preview it.
            \system offers 3 distinct methods of reconstruction, present in industry CAD software: extrude, lathe, and triplanar.
            All methods are applicable to 2D outlines or profiles made with the bricks.
            Triplanar and extrude are additionally applicable to ``solid'' tangible models (i.e. if the user has built an entire object with the bricks).
            Creating 2D outlines allows users to focus on \emph{relevant} features and generally performs better with respect to the reconstruction, as little to no occlusion is present.
            We embrace this 2D-to-3D workflow, as it allows users to explore and solve partial problems separately and combine them into a finished 3D design.
            We chose the 3 methods for a set of reasons: 1) they create useful geometry from 2D profiles, allowing users to solve partial and testable issues in 2-2.5D; 2) they are present in most, if not all, established toolchains; 3) they save users the effort of defining every single portion of the artifact, but rather ``generate'' geometry.
            
            \begin{figure}[ht]
                \centering
                \includegraphics[width=\linewidth]{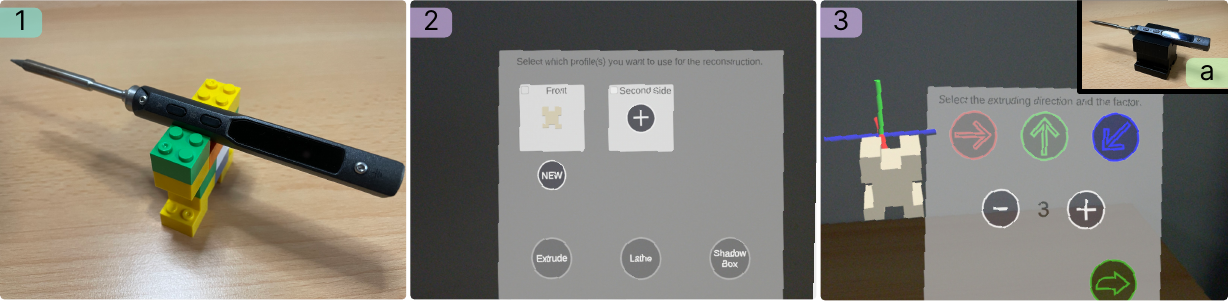}
                \caption{Applying the \emph{extrude} reconstruction method to a brick profile. The user has built a side profile of a stand for a soldering iron (1), which is photographed and reconstructed into a profile (2). Based on this profile, a 3D mesh can be generated. The user may choose an extrusion depth (3), and preview this result in-situ or fabricate it (3a).}
                \Description[]{Figure 4.: Three photos labeled 1), 2) and 3) of a brick model and interior AR screenshots from BrickStARt. Figure 4.1: Image 1) shows a brick model for a screwdriver on which one is placed. Figure 4.2: Image 2) shows on the top left the generated geometry of the model is shown. Right of it a plus is presented to add a new side. At the bottom there are three buttons Extrude, Lathe, and Shadow Box. Figure 4.3: On the left side of Image 3) the model is shown. On the right side there are three arrows (top, left, bottom left) to select the extrusion direction. Below there is a minus and plus to extrude the object in the selected direction. To go to the next step there is an arrow in the bottom right corner. On the Top right the 3D printed result of the object is shown with a screwdriver. }
                \label{fig:extrude}
            \end{figure}
            
            The \emph{extrude} reconstruction method uses a single 2D profile or side view and adds an arbitrary depth chosen by the user to it.
            This approach is common in established software, where features of a 2D sketch are selected and extruded to create a volume (e.g., \texttt{linear\_extrude()} in OpenSCAD~\cite{wikibooks2D3DExtrusion2022} or ``Extrude'' in Autodesk Fusion360~\cite{autodeskinc.ExtrudeReferenceFusion2022}).
            The process can be seen in figure \ref{fig:extrude}.

            \begin{figure}[ht]
                \centering
                \includegraphics[width=\linewidth]{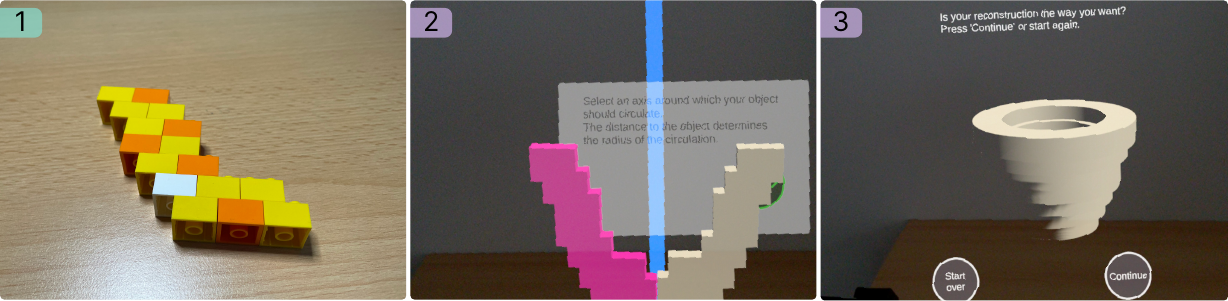}
                \caption{Applying the \emph{lathe} reconstruction method to a brick profile. The user has built a side profile of a vase (1), which is photographed and reconstructed to a 3D mesh, for which the user may choose an axis the profile is rotated around (2), and preview this result in-situ (3).}
                \Description[]{Figure 5.: Three photos labeled 1), 2) and 3) of a brick model and interior AR screenshots from BrickStARt. Figure 5.1: Image 1) shows a brick model of a side-profile of a vase. Figure 5.2: Image 2) shows the generated geometry in white. A blue line is added to the left side of this geometry. On the other side of the line, the geometry is mirrored and added in pink. The background shows a text with an instruction and a next button. Figure 5.3: Image 3) shows the generated object, which was created from the original geometry. It looks like a round vase. Below there are two buttons Start Over and Continue. }
                \label{fig:lathe}
            \end{figure}
            
            The \emph{lathe} reconstruction method also uses a single 2D profile or a side view and rotates it around an axis to create a volume.
            It is likewise present in known 3D-modeling software  (e.g., \texttt{rotate\_extrude()} in OpenSCAD~\cite{wikibooks2D3DExtrusion2022} or ``Revolve'' in Autodesk Inventor~\cite{autodeskinc.CreateRevolvedFeatures2021}).
            This process, as implemented in \system, can be seen in figure \ref{fig:lathe}.
     
\begin{figure}[ht]
                \centering \includegraphics[width=\linewidth]{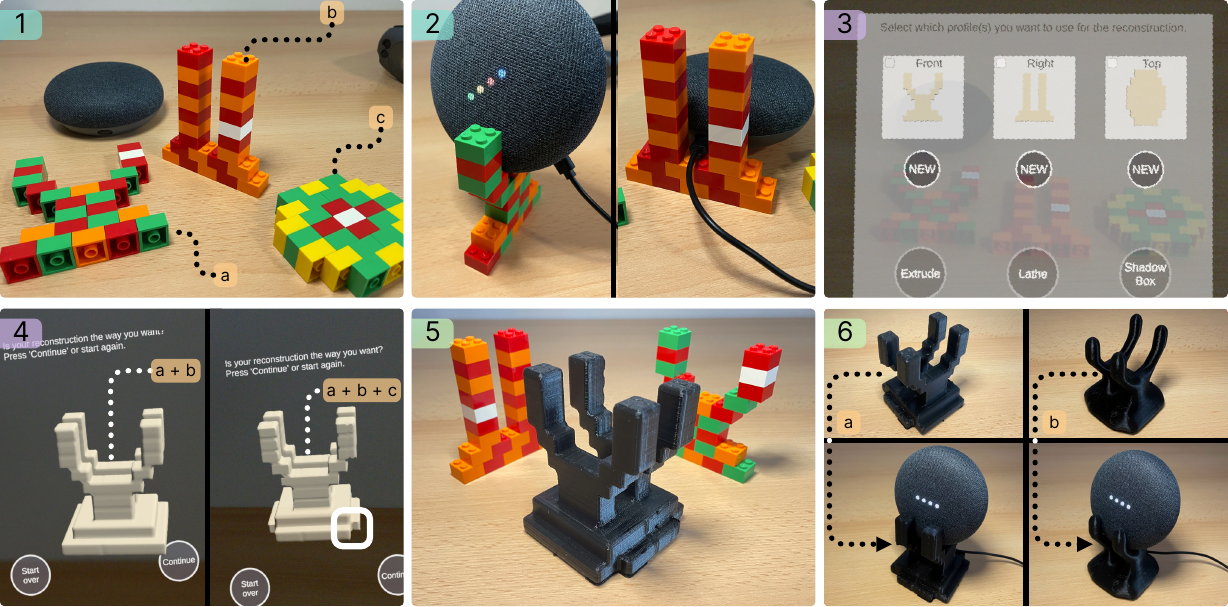}
                \caption{Applying the \emph{triplanar} reconstruction method to a set of brick profiles. The user has built 3 profiles, depicting a stand for a smart speaker from three axes (1). Profile ``a'' represents the main holding geometry, profile ``b'' is an opening for the wire, and profile ``c'' represents the footprint of the setup (2). All are acquired with \system (3). Combining 2 profiles already yields a usable result; adding the third one refines the outline further (4). The printed result mimics the initial profiles (5) and holds the speaker without (6a) and with postprocessing (6b), as intended.}
                \Description[]{Figure 6.: Six photos labeled 1), 2), 3), 4), 5) and 6) of the process to generate a stand for a smart speaker with the three planar construction method. Figure 6.1: Image 1) shows the smart speaker and three brick models a, b, and c. Model a is to hold the speaker. Model b gives the possibility to pass the cable. Model c is the stand for the construct. Figure 6.2: Image 2) is divided into two pictures. In the left picture, the smart speaker is mounted on model a and the cable goes away to the right. The right picture shows how the cable passes model b between its two columns. Figure 6.3: Image 3) is a screenshot of the BrickStARt application after generating the 3D geometry of the brick models. On the top left, there is the generated geometry of model a, on the top middle of model b, and on the top right of model c. Under each, there is a button New to restart with this geometry. On the bottom, there are the buttons Extrude, Lathe, and Shadow Box. Figure 6.4: Image 4) is a screenshot of the BrickStARt application and is divided into two pictures. The left picture shows the Shadow Box result generated of models a and b, as labeled, which is already usable as a smart speaker stand. The right picture shows the Shadow Box result generated of all three models. Adding model c causes the base of the stand to have multiple corners, making it appear more rounded. Figure 6.5: Image 5) shows the 3D printed smart speaker stand. in the background on the left there is model b and on the right model a. These show that the result mimics the initial profiles. Figure 6.6: Image 6) is divided into four parts. The top left shows the 3D printed smart speaker stand without any modifications. Below the smart speaker is mounted into it. The top right part shows the same 3D print but smoothed. Below the smart speaker is mounted into it. }
                \label{fig:triplanar}
            \end{figure}

            Lastly, \system offers users the \emph{triplanar} reconstruction alternative.
            It is likewise inspired by functions offered by screen-based modeling tools 
            (e.g., ``Triplanar'' in Nomad Sculpt~\cite{ginierTriplanarSceneNomad2022} or ``ShadowBox'' in ZBrush~\cite{pixologicinc.ShadowBoxZBrushDocs2021}), but is a rarer occurrence\footnote{openSCAD offers an example of this approach, but no built-in function: \url{https://files.openscad.org/examples/Advanced/GEB.html}, Accessed: 1.5.23} than \emph{lathe} or \emph{extrude}.
            It is present in tools focused on \emph{sculpting} (i.e., more applicable to organic shapes), and less on ``hard-surface modeling'', and is used for early definition of shapes.
            2-3 profiles or side views are placed on orthogonal planes, extruded, and intersected to yield a 3D object.
            While it is not a complete 3D scan of an object, it creates a fitting representation based on profiles or side views.
            2 profiles (e.g., top and side view) generally suffice to define a complex 3D object.
            The process can be seen in figure \ref{fig:triplanar}.

        \subsubsection{Modeling and Previewing}
            After having reconstructed the desired model digitally, users can start manipulating it in AR.
            This offers further opportunities to 'leave' the initial fixed grid of tangible building blocks.
            Users are able to freely scale the digital model.
            They can likewise rotate and move it, which allows them to preview the (altered) model in-situ, to verify aesthetic fit within the physical context, and visually estimate functional interaction~\cite{stemasovMixMatchOmitting2020}.
        
            \system also provides functionality for geometric primitives.
            Users can choose and add from a set of geometric primitives (e.g., cubes or spheres).
            These primitives can be merged with the reconstructed model to compensate for features users may have missed with their tangible model or features they were unable to model with the tangible bricks.
            This also helps users who want to circumvent the fixed grid and scale specified by the bricks initially.
        
        \subsubsection{Postprocessing}
            \begin{figure}[ht]
                \centering \includegraphics[width=\linewidth]{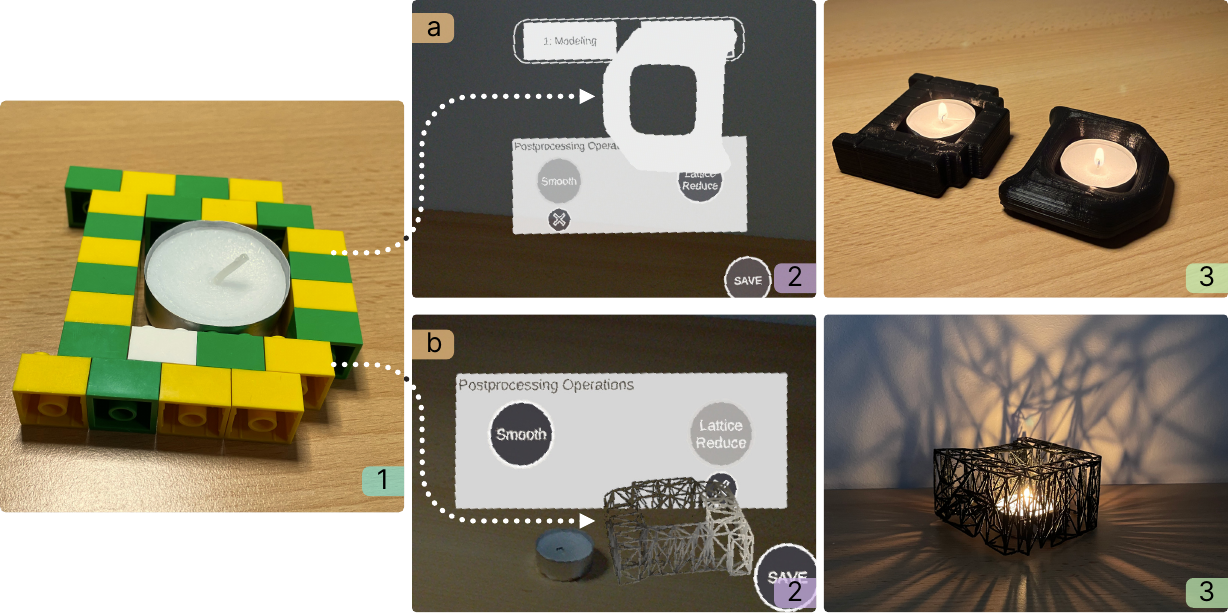}
                \caption{Postprocessing options for a candle holder: uniform smoothing~(a) and conversion to a lattice~(b). 1) original brick design with candle; 2) digital, in-situ preview; 3) fabricated results.}
                \Description[]{Figure 7.: Five photos labeled 1), a2), a3), b2), and b3) showing the several results of one brick model. Figure 7.1: Image 1) shows a brick model. Inside the model, there is a tealight. Figure 7.a2: Image a2) shows a screenshot of the application with the resulting geometry using the Lattice Reduce to generate holes inside the model. In the real environment, there is the tealight. Figure 7.a3: Image a3) shows the printed 3D model with Lattice Reduce. Inside there is a burning tealight. As a result, there are shadows on the ground and wall generated by the connections of the model. Figure 7.b2: Image b2) shows a screenshot of the application with the resulting geometry using the Smooth operation to flatten the edges. Figure 7.b3: Image b3) shows the 3D printed results. On the left side without any Postprocessing operations, on the right side the smoothed result. Both include a burning tealight. }
                \label{fig:smooth-and-lattice}
            \end{figure}
            
            Lastly, \system offers users 2 ways to postprocess their models: smoothing and a conversion to a lattice structure (fig. \ref{fig:smooth-and-lattice}).
            To match other objects in the users' environments, \system allows users to smooth the model they have made and scanned (fig. \ref{fig:smooth-and-lattice}a).
            This compensates for the blocky look (cf.,~\cite{lafreniereBlockstoCADCrossApplicationBridge2018}) introduced by the tangible building blocks, smoothing out meshing artifacts, too.
            We opted for a low degree of user control to simplify this step, applying uniform smoothing across the entire model while retaining the most distinctive features.
            Despite this approach, the benefits of the tangible building blocks remain: they may still be a tangible worst-case estimate of fit or clearance.
            Figure \ref{fig:smooth-and-lattice} depicts a candle holder made with the tangible building blocks (fig. \ref{fig:smooth-and-lattice}.1).
            The candle fits inside the cavity left open for it.
            After smoothing, this 'fit' remains (fig. \ref{fig:smooth-and-lattice}b.3), albeit with slightly more clearance.
            This is also the case for the holder where lattice postprocessing was applied to yield interesting shadow patterns (fig. \ref{fig:smooth-and-lattice}a.3).
            
            \system also offers users the possibility to convert the finished model to a lattice-like structure.
            The result of this operation can be seen in figure \ref{fig:smooth-and-lattice}a.
            We consider this to be an additional, mostly visual postprocessing step.
            However, it may also offer more functional benefits: it is possible to use the resulting structure to hold objects (e.g., as a pencil holder or a trellis for plants).
            Furthermore, this step may save material for additive manufacturing (if printed support-free), and potentially accelerate the fabrication process, as demonstrated by Mueller et al.~\cite{muellerWirePrint3DPrinted2014}.
            Lastly, users may choose to save their finalized model as an \texttt{.obj} file. 
            

%% file: sections/evaluation.tex
\section{Evaluation}
    To confirm the feasibility of \system and to explore how users may interact with it, we conducted two user studies in total: an initial pilot study with 6 participants, and a more in-depth evaluation with 16 participants.
    We furthermore evaluated \system through walkthroughs, to further probe and demonstrate its capabilities and limits.
    The following sections describe these evaluations.
    
    \subsection{Pilot User Study}\label{sec:prestudy}
        We first conducted a smaller-scale pilot study with \system.
        This pilot study was meant to elicit feedback on the concept and detect initial issues with the tool.
        
        \paragraph{Sample and Procedure}
            We recruited a set of participants relying on convenience sampling~\cite{etikanComparisonConvenienceSampling2016,henryPracticalSampling1990}. 
            Out of the 6 participants in the pilot study, 2 identified as female, and the remaining 4 participants identified as male.
            Their age ranged from 24 to 31 ($\bar{x} = 27.67$). 
            All participants were students and had at least some experience with 3D modeling tools, all of them had used interlocking building block toys before (e.g., LEGO\textregistered), and 5 of 6 had used an AR headset before. 
            During the study, participants had to solve 2 tasks.
            Task 1 involved creating a vase with \system, to familiarize themselves with the tool. 
            The second task was meant to be more challenging or at least offer more room for complexity: 
            Half of the participants were tasked with modeling a bookend; the other half had to create a coat hook with two separate hooks.
            
        \paragraph{Results}
            \begin{figure}[t!]
                \centering
                \includegraphics[width=\linewidth]{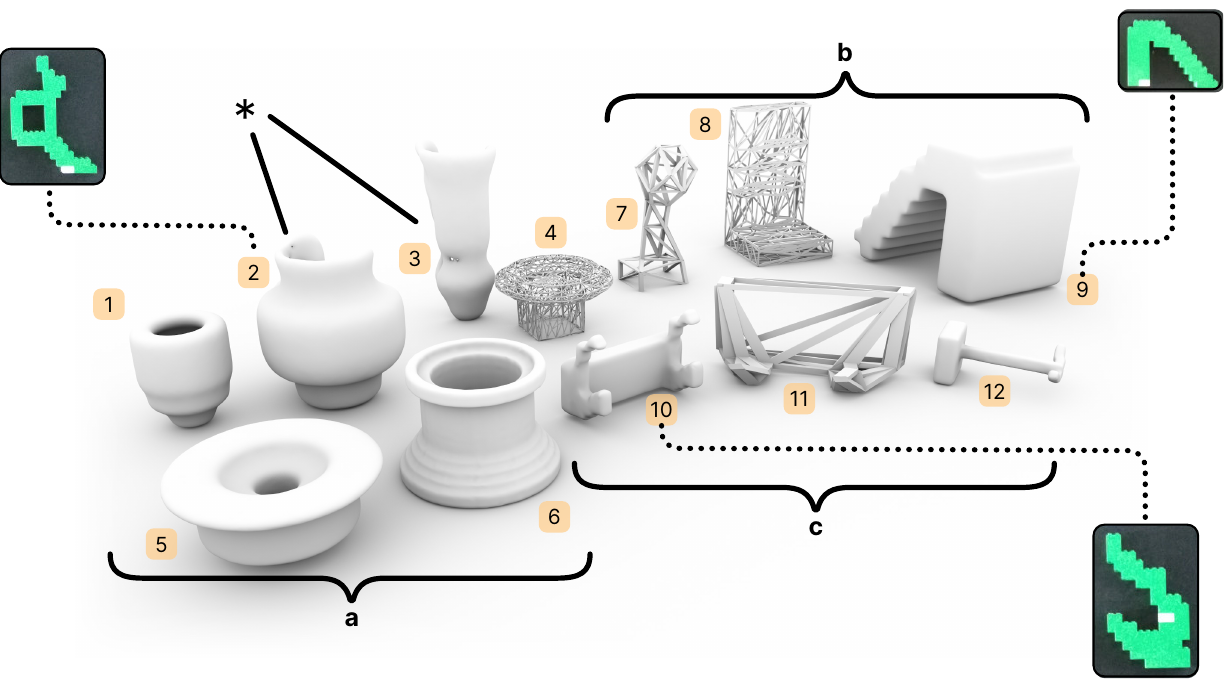}
                \caption{Results of the 6 users built during the first study: 6 vases (a), 3 bookends (b), and 3 coat hooks (c). The 2 models denoted with an asterisk (*) exhibit a faulty mesh from the reconstruction process. Select 2D brick inputs are depicted in green.}
                \Description[]{Figure 8.: Twelve model results are shown labeled from one to twelve. One to six are vases characterized as a). Seven to nine are bookstands characterized as b). Ten to twelve are coat hooks characterized as c). For models two, nine, and ten, the brick models are shown. Two and three are denoted with an asterisk exhibit a faulty mesh from the reconstruction process. }
                \label{fig:results-prestudy}
            \end{figure}
            
            The 12 resulting objects are depicted in figure \ref{fig:results-prestudy}.
            They consist of 6 vases (fig. \ref{fig:results-prestudy}a), 3 bookends (fig. \ref{fig:results-prestudy}b), and 3 coat hooks (fig. \ref{fig:results-prestudy}c).
            We found that users were able to quickly grasp \system and use it to express both functional and, to a degree, aesthetic requirements.
            \system was rated with an average score of 2.67 for the PSSUQ questionnaire~\cite{lewisPsychometricEvaluationPoststudy1992} and reached an average SUS score of 68.33.
            This was both observed by the experimenter and reflected in the comments made by the participants.
            Furthermore, the use of single-size bricks brought an unintentional design constraint, as users were forced to build their models in staggered layers.
            This is, for example, visible in \autoref{fig:results-prestudy}, models 9 and 10 and their respective brick profiles made by the participants.
            Lastly, we found the triplanar reconstruction appreciated by participants, but rarely used for the tasks given to them, despite it being applicable to both the bookend and the coat hook tasks.
            In the terminology of Resnick et al.~\cite{resnickScratchProgrammingAll2009}, this indicates a higher skill floor than intended despite offering a potentially broad range of possible geometries (wide walls~\cite{resnickScratchProgrammingAll2009}).
            In turn, the lathe and extrude functions were well-received and required little explanation, hinting at a low skill floor~\cite{resnickScratchProgrammingAll2009}.
            
            \paragraph{Takeaways for Main Study} The core limitations of the pre-study were 1) the high amount of experience of the participants with respect to CAD software; 2) the absence of a baseline to compare \system with, and 3) the fact that it focused on objects that interact with their environment physically, but these counterparts (e.g., \textbf{plant} $\Leftrightarrow$ vase) were not part of the setup and task.
            We furthermore changed our approach to the provided bricks: 
            in the second study, users were encouraged to use multicolored brick assemblies to support recognition, and we added longer bricks to the users' selection to reduce the influence of having only one brick size available on the design.
            The lighting in the study environment was also improved to support the algorithm further.
            In addition, a set of minor issues with the implementation (e.g., how users grab and reposition objects in space, unresponsive buttons) were rectified.
    
    \subsection{User Study}\label{sec:study}
        The follow-up study was conducted to engage more deeply with the concept of tangible, in-situ 3D modeling by actively taking into account interacting objects from the physical environment (i.e., \emph{context}) and comparing \system  with a more established, novice-friendly 3D modeling tool: Autodesk \tk\footnote{\url{https://www.tinkercad.com/}, Accessed: 2.5.23}. 
        \subsubsection{Procedure and Sample}
            \paragraph{Study Design}
                The study followed a 2x2 within-subjects design, combining the factors \texttt{tool} (levels: \tk and \system) and \texttt{task} (levels: phone stand and vase), with participants designing one model with each tool.
                The order of tasks and tool combinations was counterbalanced, and we ensured an even distribution of combinations and orders using a Latin Square to avoid carryover effects.
                Following the insights from the pilot study, we chose the 2 tasks based on the following criteria: 1) meaningful physical interaction with the environment (i.e., the phone and the potted plant), 2) simple enough geometry to be designed within 10-15 minutes, and 3) the potential for aesthetic expression of the participants.
                When observing the most popular designs on Thingiverse\footnote{\url{https://www.thingiverse.com/}, Accessed 2.5.23}, phone/tablet stands and vases are a reasonably common type of design (in addition to mechanical parts, organization tools, and various figurines).
                For the alternative tool, we settled on \tk, established in the personal fabrication community as a particularly accessible 3D-design tool~\cite{bhaduri3DnSTFrameworkUnderstanding2021,mueller3DPrintingHumancomputer2017}.
                Novices had to quickly learn 2 new tools over the course of the study, and both \tk and \system can, arguably, be considered suitable for that.
                
            \paragraph{Metrics} 
                We acquired a set of qualitative and quantitative metrics (i.e., dependent variables) for the evaluation.
                Primarily, we were interested in users' success rates and the specific errors they may make in case of failure.
                We formalize ``success'' primarily based on function: users were tasked with ``making a phone stand'' and ``making a vase''. 
                To further guide the users, each task consisted of 2 core objectives and an optional one.
                Users were handed printed cards that summarized the objectives and could reference them anytime.
                The phone stand task asked for ``an angle of approximately 50\textdegree-60\textdegree'', the possibility for it to ``hold the phone in vertical and horizontal orientation'', and, as an optional objective, ``allow a charging cable to pass through''.	
                The vase task asked for ``less than 1cm of play'' between the inner and outer pot, a design that does not ``topple easily'', and optionally, a visually interesting feature on the vase.
                These aforementioned criteria can be formalized under the functionality of the object.
                In addition to that, the stability (i.e., balance) and the fit (i.e., with respect to the interacting object) were relevant rating dimensions. 
                The time participants took to ``solve'' (following their own judgment) a task was recorded.
                We further recorded the raw NASA TLX (Task Load Index)~\cite{hartNASAtaskLoadIndex2006} and the SUS (System Usability Scale)~\cite{brookeSUSAQuickDirty1996} after each task, while also presenting sets of non-standardized statements to be rated on Likert-Scales.
                Lastly, we solicited comments from all participants, specifically asking for positive and negative aspects of the respective tools.
                To better understand users' processes, their actions were recorded on video after explicit consent -- \tk tasks were recorded using OBS\footnote{Open Broadcaster Software, \url{https://obsproject.com/}, Accessed: 1.5.23} on a laptop computer and \system tasks were recorded from a smartphone pointed at the participant's working area (a table).
            
            \paragraph{Participants}
                We recruited 16 participants, across a broader range of users than in our pilot study.
                Out of the 16 participants, 7 identified as female, and 9 participants identified as male.
                Their age ranged from 18 to 33 ($\bar{x} = 23.63$).
                5 participants have used 3D modeling software before, and all 16 had experience with building blocks.
                9 had used an AR headset before, but did so for a total duration below 10 hours (e.g., in the context of other user studies).
                The participants were mostly students with various backgrounds (e.g., computer science, biology, finance, mathematics).

            \paragraph{Procedure}
                A general introduction regarding the study procedure was done first.
                Participants had to do 2 tasks in total: a phone stand and a vase, either using \system or \tk.
                Prior to each task, participants received a brief introduction to the tool and the criteria relevant to their task.
                Similarly, they received a brief introduction to 3D-modeling for fabrication as such.
                For \system, they were introduced to the process (outlined in \autoref{fig:process}), as implemented in the system, first.
                They were then introduced to fundamental interaction techniques for the HMD (i.e., controller inputs and gestures) and the general interaction flow of \system.
                For \tk, users received a similar introduction involving all fundamental functions needed to achieve their goals (e.g., object manipulation, translations, primitives).
                Similarities between the tools were mentioned (e.g., the presence of Boolean operations) where fitting.
                For both introductions, users were allowed and encouraged to use and explore the system for up to 5 minutes, familiarizing themselves with the respective tool.
                Users were allowed to ask questions about functions at this stage.
                They were informed that their design was meant to be 3D-printed, and would be made out of plastic.
                Interacting objects were available to the users: a phone (LG Nexus 5X, $147 \times 73 \times 7.9 mm$) and a small houseplant ($D_1=50mm, D_2=38mm, h=50mm$).
                For the \tk tasks, users were also given a caliper and a ruler to measure these interacting objects.
                Participants were suggested to take 15 minutes to complete their design, but there was no hard cutoff after these 15 minutes, letting all participants finalize their design.
                On average, the study took between 50 and 70 minutes.
                The tasks themselves took between 13 and 32 minutes using \system\footnote{the longer durations are partly explained by technical issues, but also due to some users strongly focusing on their brick model} and between 7 and 15 minutes using \tk.
                All participants were reimbursed 11~\texteuro~for their time.

        \subsubsection{Results}
            \begin{figure*}[h]
                \centering \includegraphics[width=\textwidth]{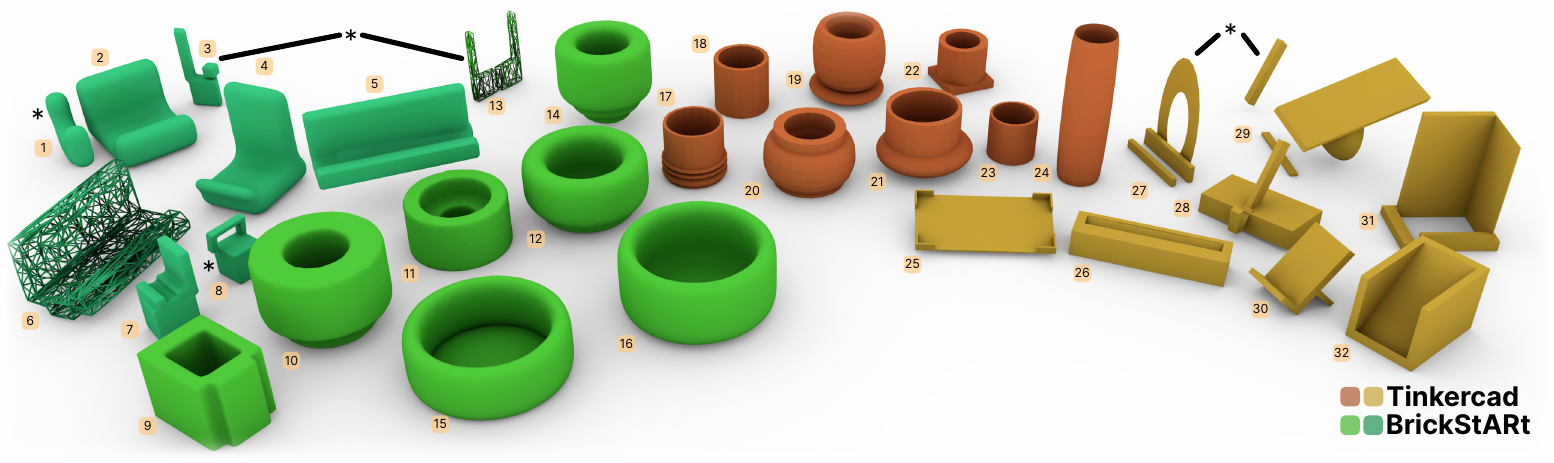}
                \caption{Models made by the participants in the second user study: vases made with \system (light green) and \tk (orange), and phone stands made with \system (teal) and \tk (yellow). Models denoted with an asterisk (*) exhibit a faulty mesh (i.e., one that might be manufacturable, but would not function due to its geometry) from the scanning process (in \system) or from design errors (in \tk).}
                \Description[]{Figure 9.: 32 model results are shown labeled from 1 to 32. 1 to 8 are phone stands made with BrickStARt and are teal-colored. 9 to 16 are vases made with BrickStARt and are colored light green. 17 to 24 are vases made with Tinkercad, colored orange. 25 to 32 are phone stands made with Tinkercad, colored yellow. Models 1, 3, 8, 13, 27, and 29 are marked with an asterisk and exhibit faulty meshes. }
                \label{fig:results2}
            \end{figure*}
        The models made by the participants are depicted in figure \ref{fig:results2} -- 16 vases and 16 phone stands.
        Out of the 32 models made, 11 models completely fulfilled their intended function, 15 did so at least partially, and 6 did not.
        When split by the tool used, models made with \tk were non-functional 2 times, partially functional 9 times, and functional 5 times.
        In contrast, models made with \system were non-functional 4 times, partially functional 6 times, and functional 6 times.
        Completely non-functional results emerged either through issues during scanning with \system (fig. \ref{fig:results2}: 1, 3, 8, 13), or through errors made by the users while modeling with \tk (fig. \ref{fig:results2}: 27, 29).
        
        \paragraph{Error Analysis:}
            \begin{figure}[hb]
                \centering
                \includegraphics[width=\linewidth]{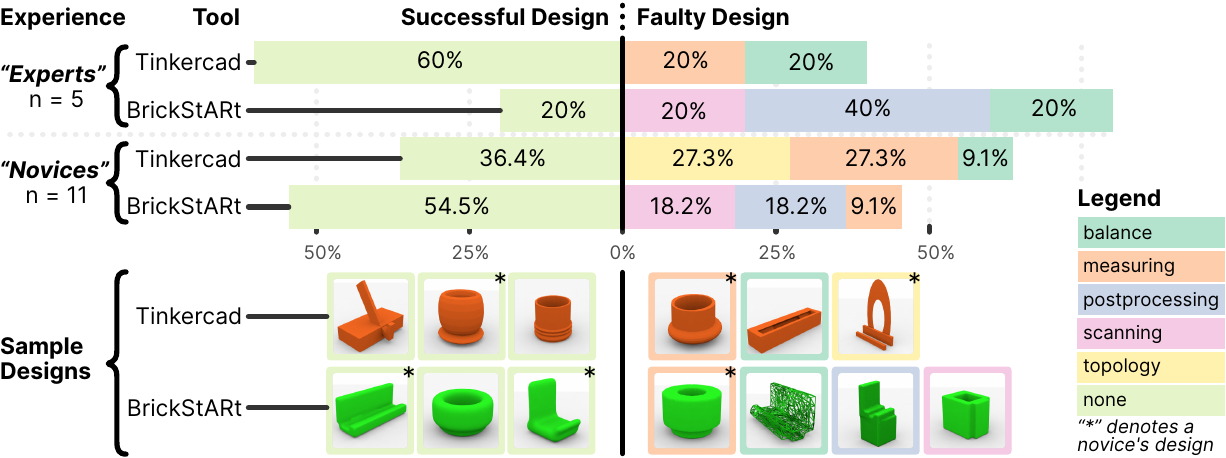}
                \caption{Core design issues split by prior CAD experience (column 1) and tool used (column 2). Faulty designs are further classified by their core issue (color coding, see legend). Below: samples of designs and design issues made with the tools, first row -- \tk, second row -- \system (outline color depicts core issue of design, asterisks denote a novice user).}
                \Description[]{Figure 10.: This figure contains a horizontal stacked barplot. It has an x-axis with labels 0\%, 25\%, 50\%, 75\% and 100\%. It has a y-axis with 2 groups, each separated in 2 more groups. The first grouping in the leftmost column of the plot, it refers to CAD experience, a green checkmark is in the first cell, with n=5 below it. A blue cross is in the cell below, with n=11 written below it. The second column of the grouping is the 'tool used'. Each cell from the fist column is now split into two in this second column. In this second column, TK, BST, TK and BST are labelled in this order, each to the left of a stacked barplot. There is a legend indicating fill is used to show the issue, with 6 levels: balance shown as very pale green fill, measuring shown as pale orange yellow fill, postprocessing shown as very pale purplish blue fill, scanning shown as pale purplish pink fill, topology shown as pale yellow green fill and none shown as pale yellow fill. The first stacked bar refers to CAD Experience: yes, and Tool Used: TK. The first bar part of this bar is 'none' and is labelled with 60\%. It is colored in a pale green fill. The second bar part of this bar is 'measuring' and is labelled with 20\%. It is colored in a pale orange yellow fill. The third bar part of this bar is 'balance' and is labelled with 20\%. It is colored in a pale fill. The second stacked bar refers to CAD Experience: yes, and Tool Used: BST. The first bar part of this bar is 'none' and is labelled with 20\%. It is colored in a pale green fill. The second bar part of this bar is 'scanning' and is labelled with 20\%. It is colored in a pale purplish pink fill. The third bar part of this bar is 'postprocessing' and is labelled with 40\%. It is colored in a pale purplish blue fill. The fourth bar part of this bar is 'balance' and is labelled with 20\%. It is colored in a pale fill. The third stacked bar refers to CAD Experience: no, and Tool Used: TK. The first bar part of this bar is 'none' and is labelled with 36.4\%. It is colored in a pale green fill. The second bar part of this bar is 'topology' and is labelled with 27.3\%. It is colored in a pale yellow green fill. The third bar part of this bar is 'measuring' and is labelled with 27.3\%. It is colored in a pale orange yellow fill. The fourth bar part of this bar is 'balance' and is labelled with 9.1\%. It is colored in a pale fill. The first stacked bar refers to CAD Experience: no, and Tool Used: BST. The first bar part of this bar is 'none' and is labelled with 54.5\%. It is colored in a pale green fill. The second bar part of this bar is 'scanning' and is labelled with 18.2\%. It is colored in a pale purplish pink fill. The third bar part of this bar is 'postprocessing' and is labelled with 18.2\%. It is colored in a pale purplish blue fill. The fourth bar part of this bar is 'measuring' and is labelled with 9.1\%. It is colored in a pale orange yellow fill.  }
                \label{fig:issues}
            \end{figure}
            
            With the finished designs, we proceeded to conduct a more in-depth analysis of errors and issues in the models.
            3 core categories were used to evaluate the designs: \textbf{fit} (whether the plant or phone fit), \textbf{balance} (whether the object would stand on its own), and \textbf{function} (whether it fulfills the core function as intended).
            Each design was scored with respect to the 3 categories on a discrete scale of ``yes'', ``partial'', ``no''.
            Designs with deficiencies were then labeled by 2 of the authors to determine the core issue of a design.
            Comments regarding the function and errors were gathered initially and then converted to labels representing recurring themes.
            For non-obvious issues (i.e., ones where one would either have to set up a simulation to verify them, or fabricate them), or disagreements between the 2 raters, the models were 3D printed with Cura's default settings in PLA filament and tested manually with the interacting objects users relied on in the study (e.g., fig. \ref{fig:results2}: 7, 11, 24, 26, 30).
            While evaluating failure reasons, we formalized 5 recurrent reasons for the users' designs: balance, measuring, topology, postprocessing, and scanning (the latter 2 being specific to \system).
            The distribution of issues is depicted in figure \ref{fig:issues}.
            \textbf{Balance} issues were rated for objects that are unable to stand on their own (e.g., fig. \ref{fig:results2}: 6, 30), or would be easy to topple using a light touch (e.g., fig. \ref{fig:results2}: 14).
            \textbf{Measuring} issues arose when users either measured incorrect values, or transferred them incorrectly to the design environment, leading to incorrect clearances (e.g., fig. \ref{fig:results2}: 25, 32).
            \textbf{Postprocessing} issues were unique to \system and describe the digital alteration of a design that may have been functional, but was rendered unusable after scaling (e.g., fig. \ref{fig:results2}: 7), or applying other postprocessing operations (e.g., fig. \ref{fig:results2}: 14).
            \textbf{Scanning} issues were similarly unique to \system, and describe a class of problems where \system was unable to reliably reconstruct the intended design from the scanned brick model (e.g., fig. \ref{fig:results2}:1, 3, 8, 13).
            \textbf{Topology} issues describe the design of non-functional or non-practical geometries, for instance, are missing connecting elements to work (e.g., fig. \ref{fig:results2}: 27, 29).
            
            For the \textbf{overall function}, the success rate was comparable ($BST:7,TK:7$), while partial functionality was more often achieved with \tk ($BST:2,TK:4$).
            For \textbf{stability}, a similar picture emerged, with \tk performing slightly better (stable: $BST:9,TK:10$, unstable $BST:3,TK:2$).
            This was also the case for \textbf{fit} (fits: $BST:4,TK:4$, does not: $BST:7,TK:8$).
            While the overall failure rate of both systems was fairly similar, we conducted a deeper labeling of the different \emph{types} of errors. 
            We found that the majority of errors in \system occurred due to current implementation issues (e.g., scanning 18.8\% or postprocessing 25\%), and that balance or clearance issues were a rarer occurrence in \system, compared to \tk ($BST:12.4\%,TK:37.5$).
            This has, however, to be considered under the assumption that the system failures in \system would have yielded impeccable designs, which is not necessarily the case.
            Additionally, 3 designs made with \tk failed due to incorrect topology (i.e., disconnected parts -- fig. \ref{fig:results2}: 27, 29, 31).
            
            \textbf{Prior experience} with 3D modeling seemed to have an effect on which errors users made and whether they made any at all (fig. \ref{fig:issues}).
            With no prior experience (i.e., \textit{novices}), 6 users succeeded with \system, compared to 4 successes with \tk.
            Only 1 of the novices made measurement/clearance errors with \system, compared to 3 using \tk, and no novice using \system misjudged balance.
            Novices using \tk also made errors in topology, forgetting to connect elements of their models, which was not the case for \system.
            In turn, users with prior experience generally performed better with \tk, with 1 occurrence of measurement error and 1 balance issue.
            With \system, 3 failures were due to technical issues, and 1 due to misjudged balance, emerging after smoothing.
            
        \paragraph{Questionnaire Data: }
            All participants agreed that using the bricks was helpful for their design process and generally agreed that with both tools, they were able to estimate sizes ($BST: 62\%, TK: 88\%$) and proportions ($BST: 75\%, TK: 69\%$). 
            Similarly, a majority of participants (81\%) agreed that both \tk and \system allowed them to explore how their object would look like after fabrication.
            \tk achieved a higher average SUS score than \system (75.6 vs. 53.6), which primarily originated from technical issues, such as imprecise reconstruction and errors in postprocessing leading to frustration.
            This explanation was further supported by the comments left by the participants.
            44\% of users would have liked to 3D-print their design after making it with \system, compared to 35\% of the \tk users.
            In contrast, only 12\% of the users rated their design as ``pleasing'' for \system, compared to 38\% for \tk. 

            When asked for which tasks they would prefer either of the 2 tools (on a 5-point-scale), users preferred \system for objects where they want to understand physics ($BST:56\%, TK:12\%$), objects that do not require precise measurements ($BST:56\%, TK:19\%$) and objects for which they may not immediately know a design approach ($BST:44\%, TK:31\%$).
            \tk, in contrast, was preferred for objects demanding precision ($BST:0\%, TK:88\%$), objects where the solution is known ($BST:19\%, TK:75\%$), and complex objects ($BST:19\%, TK:75\%$).
            These patterns changed slightly when observing novices compared to experienced users:
            60\% of users with prior experience leaned towards \system for which they do not immediately know a solution to, and 80\% of them leaned towards \system for objects that do not require precise measurements. 
            Prior experience with AR had an effect on tool preference regarding the \emph{looks} of objects. 
            For ``objects for which looks matter'', users with prior AR experience preferred \system ($BST:78\%, TK:22\%$), while users without prior AR experience leaned towards \tk ($BST:14\%, TK:71\%$). 
            This pattern repeated for the contrary of the question (``objects for which looks do not matter''): users with prior AR experience preferred \tk ($BST:22\%, TK:67\%$), while users without prior AR experience leaned towards \system ($BST:57\%, TK:14\%$). 
            The raw NASA TLX exhibited little difference between the tools, except for the physical demand and the temporal demand scales.
            \system exhibited a higher physical demand than \tk.
            Users also generally reported a lower temporal demand while using \system than \tk.

        \paragraph{Comments and Observations: }
        \system was acclaimed for the intuitivity it had during the tangible modeling phase (\textit{``handy''} -- P2, ).
        Tangible modeling was further praised for being \textit{``... playful [and] ... making the start of modeling rather easy''} (P13).
        For all participants, we observed them testing their design with their brick model combined with interacting objects (i.e., plant, phone) at least once prior to scanning.
        This approach to defining separate features to extrude/revolve later was also appreciated (\textit{``... efficient and intuitive''} -- P5, \textit{``... practical''} -- P7).
        The AR aspects of \system were similarly well-received, as it was \textit{``very helpful for roughly estimating sizes''} (P10),
        \system was \textit{``more visual''} (P8) and that \textit{``... the visualization helped embed the size of the phone in the process''} (P16).
        These benefits echo observations during the procedure, where users actively embedded the environment and interacting objects in their design process, leveraging the spatial nature of the interface and process during the AR phase, \emph{continuously interacting in-situ}.
        Negative comments referring to \system were primarily due to issues in scanning (P1, P8, P11) or during postprocessing (P11, P13), which have affected user experience negatively.
        Users also were missing features for fine-grained manipulation (P5, P10).
        \tk was similarly well-received, as it was \textit{``Very easy to estimate sizes and proportions, easy to fix mistakes quickly''} (P3), and it provided \textit{``Precise rotation and scaling tools''} (P11).
        However, it was also criticized for view controls (P7, P10, P12, P15) and issues while combining objects or features -- \textit{``It was somewhat unclear if there are gaps or not between two objects''} (P9).
        This was relevant for several of the models (e.g., fig. \ref{fig:results2}: 19, 25, 29). 
        
        \input{sections/walkthroughs.tex}

%% file: sections/walkthroughs.tex
\subsection{Walkthroughs}\label{sec:walkthroughs}
    The following sections depict three walkthroughs through tasks solvable with \system.
    With these use-cases, we want to expand upon the user studies, by providing a more detailed view of the potential capabilities of \system, highlighting approaches that were not feasible to test in the short timeframes in either of the 2 studies.
    
    \subsubsection{Phone Stand -- In-situ Physics Exploration}
        \begin{figure}[h!b]
            \centering
            \includegraphics[width=\linewidth]{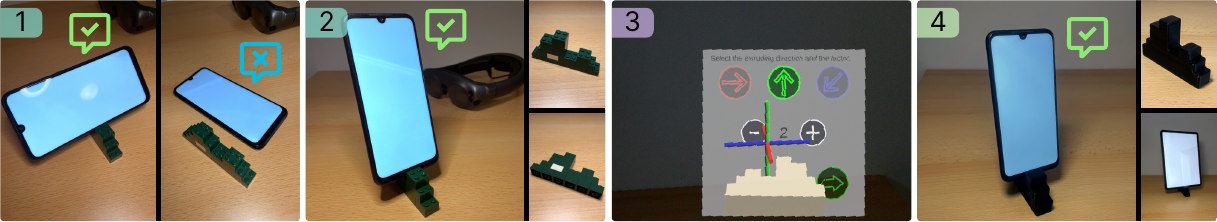}
            \caption{A user's goal is to create a phone stand that works for vertical and horizontal orientations of the device. 1) initial model, partially fulfilling the requirement; 2) further bricks are added for stabilization; 3) acquisition and extrusion of the profile; 4) fabricated result, holding the phone as expected and a tablet beyond that.}
            \Description[]{Figure 11.: Four photos labeled 1), 2), 3), and 4) adding bricks to the brick model for a phone stand to stabilize. Figure 11.1: Image 1) is divided into two parts. On the left side, there is a brick model and a phone horizontally mounted on it. A checkmark shows that the phone can be held with this brick model. On the right side, there is the brick model and the phone lays beside it. Here a cross shows that the phone cannot be mounted vertically on this model. Figure 11.2: Image 2) is divided into three parts. The main part shows the phone standing vertically on the brick model. A brick was added to enable this position for the phone. On the top right, the model is standing and on the bottom right, the model is laying on a table. Figure 11.3: Image 3) shows a screenshot of the application. The generated geometry is before a canvas to extrude the model. The canvas has three arrows are shown to extrude the object in an axis: top, right, bottom left. Below there are a plus and a minus to increase and decrease the geometry. Figure 11.4: Image 4) is divided into three parts. The main part shows the phone on the 3D printed model created of the brick model. On the top right, only the printed model is shown. On the bottom right, a tablet is vertically mounted on the printed model. }
            \label{fig:walkthrough-phonestand}
        \end{figure}
        
        \system enables users to explore physical consequences in their model early in the design process.
        Iteration then happens while using reconfigurable, reusable tangible building blocks, instead of fabricating a prototype (e.g., 3D-printing it), testing it, and resolving issues that were detected.
        In figure \ref{fig:walkthrough-phonestand}, a user wants to create a simple phone stand.
        The stand is supposed to hold the user's phone, while allowing it to be oriented vertically (for video calls) or horizontally (for watching videos), akin to the task in study 2.
        The user creates an initial model, which works for the horizontal orientation (fig. \ref{fig:walkthrough-phonestand}-1).
        After re-orienting the phone, the construction topples.
        The user then adds additional bricks (fig. \ref{fig:walkthrough-phonestand}-2).
        Notably, the user can test the model for stability in 1 axis, despite being essentially ``flat'' (i.e., one brick deep).
        The corrected and validated design is now scanned and extruded to match the phone's width (fig. \ref{fig:walkthrough-phonestand}-3).
        The resulting model can be previewed in-situ and altered further, if needed.
        The fabricated result (fig. \ref{fig:walkthrough-phonestand}-4) fulfills the initial requirements and may even hold larger devices, which indicates how the bricks serve as a ``worst-case-estimation'', albeit one that may obscure the full capabilities of the design.
    
    \subsubsection{Coat Hook -- Varying Paths to Success}
        \begin{figure}[h!]
            \centering
            \includegraphics[width=\linewidth]{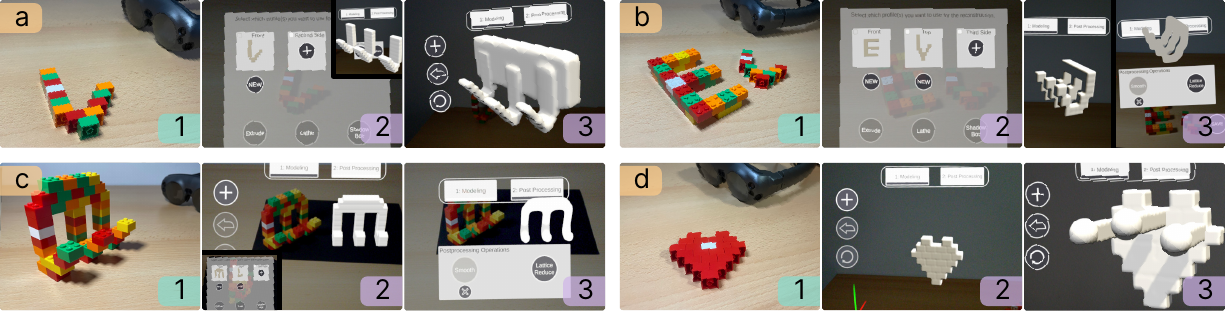}
            \caption{4 approaches to the creation of a three-pronged coat hook with \system. 
            a) multiplying an initial design; b) triplanar reconstruction from separate profiles; c) triplanar reconstruction from a solid model, smoothed in postprocessing; d) acquired backplate combined with primitives.}
            \Description[]{Figure 12.: Four image series, each three pictures, labeled a1), a2), a3), b1), b2), b3), c1), c2), c3), d1), d2), and d3), to create different coat hooks. Figure 12.a.1: Image a1) shows a single brick-built hook looking like a reversed J. Figure 12.a.2: Image a2) is a screenshot of the application and shows the scanned brick model. Below the button, NEW is to recreate the geometry. Besides, it is possible to add two new sides to the model. Below there are the buttons Extrude, Lathe, and Shadow Box. On the top right, the picture shows the hook model three times next to each other. Figure 12.a.3: Image a3) is a screenshot of the application and shows the three hooks with an added cube to connect them at the back to have one object instead of three. Figure 12.b.1: Image b1) shows two different brick models. One looks like an E and one like a reversed J. Figure 12.b.2: Image b2) is a screenshot of the application and shows the scanned brick models. Below the button, NEW is to recreate the geometry. Besides, it is possible to add one new side to the models. Below there are the buttons Extrude, Lathe, and Shadow Box. Figure 12.b.3: Image b3) is a screenshot of the application and is divided into two parts. On both sides, the Shadow Box option was applied with both generated geometries. On the left side, a coat hook is created with no further options. On the right side, the smooth function was additionally added to the generated object. Figure 12.c.1: Image c1) shows a single brick model placed on a table. The model shows a hook with 3 rounded pins protruding from it. Figure 12.c.2: Image c2) is a screenshot of the application and shows the scanned sides of the brick model. Below the button, NEW is to recreate the geometry. Besides, it is possible to add one new side to the models. Below there are the buttons Extrude, Lathe, and Shadow Box. To the right, another screenshot is visible. The brick model is seen beside its augmented reconstruction. Figure 12.c.3: Image c3) is a screenshot of the application and depicts screenshot of the application. A smoothed version of the hook from the previous images is visible, along with the UI showing that ``smoothing'' is enabled. Figure 12.d.1: Image d1) shows a heart-formed backplate of bricks for the coat hook. Figure 12.d.2: Image d2) is a screenshot of the application and shows the scanned brick model without any modification. Figure 12.d.3: Image d3) is a screenshot of the application. The backplate has added components. These components consist of a cuboid and a sphere. Both result in one hook. This result is added three times to the backplate. }
            \label{fig:walkthrough-coathook}
        \end{figure}

        Figure \ref{fig:walkthrough-coathook} depicts 4 different approaches to the creation of a three-pronged coat hook.
        The first approach (fig. \ref{fig:walkthrough-coathook}-a) allows users to focus on the most important part of the object: the hook itself.
        Users initially build a tangible model of a single hook, acquire, and extrude the scanned design.
        Afterward, users can duplicate their design, arrange the resulting hooks, add a cube to serve as a backplate, scale it to match the distance between the hooks, and combine the 3 objects.
        The second approach (fig. \ref{fig:walkthrough-coathook}-b) depicts a user creating 2 profiles instead of one.
        One depicts a single hook itself, the other describes the connecting geometry between the 3 hooks.
        As before, the outlines are acquired by \system, combined through the triplanar reconstruction, and presented to the users for in-situ previewing.
        Lastly, the users may also smooth the result, which yields a more organic shape.
        The third approach (fig. \ref{fig:walkthrough-coathook}-c) involves a user creating the \emph{entire} hook with their tangible model.  
        The model is photographed from 2 sides and, as before, the triplanar reconstruction is applied. 
        This yields a mostly faithful recreation of the original model.
        The fourth approach (fig. \ref{fig:walkthrough-coathook}-d) involves a user focusing more on the decorative geometry of the hook's backplate.
        This backplate is built first, using the tangible building blocks.
        It is acquired with \system and extruded to a default depth (i.e., one that is one brick deep).
        The users then add elongated cubes to the plate and merge them into a single model, resulting in a functional hook.
        This approach treats \system more like a primitive-based 3D-modeling tool, as the most (functionally) relevant features of the hook are created digitally only.
        
    \subsubsection{Bookend -- Dealing with Scale \& Resolution}
        \begin{figure}[h!]
            \centering
            \includegraphics[width=\linewidth]{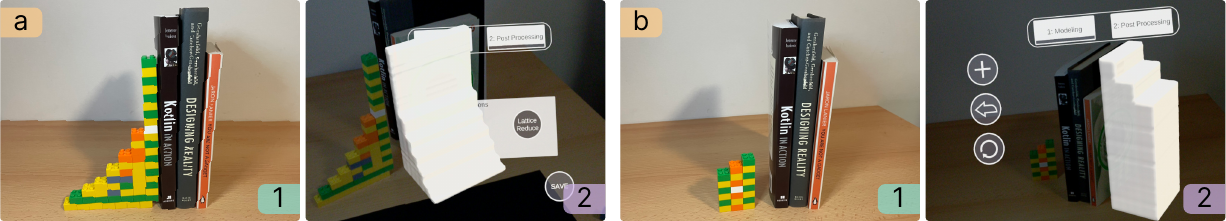}
            \caption{Creation of a bookend with \system while dealing with its low spatial resolution. Using a to-scale tangible design~(a). Creating a tangible design that is not-to-scale, and postprocessing + scaling it to fit in AR~(b).}
            \Description[]{Figure 13.: Four pictures labeled a.1, a.2, b.1, and b.2 constructing a book stand with BrickStARt. Figure 13.a.1: Image a.1 shows three books held by a bookstand built with bricks with one brick depth. Figure 13.a.2: Image a.2 shows a screenshot of the application where the geometry of the brick model is extended in depth. In the background of the model, the books of the real environment are shown to compare the geometry size with the size of the books. Figure 13.b.1: Image b.1 shows a brick model which is much smaller than the books beside it. Its depth is one brick. Figure 13.b.2: Image b.2 shows a screenshot of the application where the geometry of the brick model is extended in depth and height to fit the books. In the background of the model, the books of the real environment are shown to compare the geometry size to the size of the books. }
            \label{fig:walkthrough-bookend}
        \end{figure}
        
        The use of the tangible building blocks enforces a certain resolution and therefore also a certain angularity.
        This is appropriate and beneficial for early ideation and iteration, but may not be desirable for a finished, fabricated object.
        Using \system, users are able to compensate for this in three ways: by creating a \emph{larger} tangible model than necessary, creating a \emph{smaller}, or by smoothing a to-scale model in AR later on.
        Figure \ref{fig:walkthrough-bookend} depicts this for the task of creating a bookend.
        One option is creating a to-scale tangible model (fig. \ref{fig:walkthrough-bookend}a). 
        This enables the exploration of proportions early on. 
        Alternatively, users may create a tangible model that is not-to-scale (fig. \ref{fig:walkthrough-bookend}b).
        This requires fewer bricks, but allows users to achieve a desired shape and scale anyway.
        This smaller tangible model can be acquired by \system and scaled up to a reasonable size later on.
        This result can likewise be smoothed if necessary.

%% file: sections/discussion.tex
\section{Discussion}\label{sec:discussion}
    Building, exploring, and evaluating \system has helped us understand tensions of \emph{tangible}, \emph{in-situ}, and \emph{low-effort} 3D-modeling tools for personal fabrication.
    In the following sections, we discuss the evaluation results, followed by a broader discussion of the concept behind \system.
    
    The study confirmed that \system is not only a feasible and useful tool for simple personal fabrication tasks but also one that allows users to test and iterate designs early, without fabricating design iterations~\cite{muellerFaBrickationFast3D2014}, at the expense of fidelity or precision. 
    This positively affected the models made by participants without any prior CAD experience, who made fewer errors regarding measurements and physics: 54.5\% of the first-time 3D-modelers immediately succeeded using \system, compared to 36.4\% using \tk, with only 9.1\% of them failing to acquire the correct measures with \system.
    In contrast, participants already familiar with 3D-modeling generally performed better with \tk, as it largely follows established paradigms of solid modeling. 
    We observed several technical issues (recognition quality and robustness), which we consider relevant technical shortcomings. 
    Recognition of the bricks could likely be improved through other approaches that are not purely based on 2D vision~\cite{tattersallReconstructingCreativeLego2021}, and further leverage the HMD as a (depth) sensor platform~\cite{stemasovMixMatchOmitting2020}.

    \paragraph{Process}
        A relevant disadvantage of \system is the \textbf{phase disconnect} between tangible modeling and digital postprocessing.
        It becomes apparent when users apply postprocessing steps in the digital phases of \system (e.g., scaling, or addition of primitives).
        While postprocessing steps are beneficial, they yield an additional disconnect from the tangible blocks: aspects like balance, clearance, or size, in general, have to be estimated visually only.
        This still happens in-situ, but in an intangible fashion.
        Postprocessing steps like smoothing had little effect on previously estimated aspects of balance and fit (e.g., fig. \ref{fig:results2}: 2, 4, 15, 16).
        The addition of primitives (e.g., fig. \ref{fig:results2}: 7, which surprisingly worked anyway) or scaling (e.g., fig. \ref{fig:results2}: 14), however, did.
        In contrast, \tk exhibits this disconnect between the measurement phase and the modeling phase, where users made errors in dimensioning specific features of their designs (e.g., fig \ref{fig:results2}: 18, 25, 26, 32), which was identified as challenging in prior works~\cite{mahapatraBarriersEndUserDesigners2019}.
    
    \paragraph{Tradeoffs between \system and \tk}
        \textbf{Measurement-based errors} were more common in \tk.
        For measurements, we observed that users generally tried to be precise in \tk.
        This is possible and afforded by \tk, but gives rise to small errors that compromise the entire results (e.g., fig. \ref{fig:results2}: 25, 32).
        Having to operate with the fixed grid required by \system forced users to embed more clearance to interacting objects, where possible, and \textit{``err on the side of clearance''}, which in turn does not compromise the results (e.g., fig. \ref{fig:results2}: 2, 4, 15, 16), but does not yield tight tolerances either.
        This contrasts fundamental notions of personal fabrication, where \emph{precision} in manufacturing is handed to end-users~\cite{baudischPersonalFabrication2017,gershenfeldFabComingRevolution2005}.
        The absence of a readily accessible simulation component in \tk made it harder to detect balance issues early on\footnote{After our study, Tinkercad independently added a simulation environment to the software: \url{https://www.tinkercad.com/blog/tinkercad-sim-lab}, Accessed 22.7.2023 }. 
        Users would have had to fabricate prototypes or set up a simulation elsewhere to detect them. 
        This lead to designs that, at first glance, seem functional, but fail when fabricated and tested (e.g., fig. \ref{fig:results2}: 19, 24, 26, 30).
        \textbf{Tangible testing and exploration} is at the core of \system.
        Users could detect and compensate for some issues referring to clearance and balance (without simulation) with little to no explicit measuring effort.
        Simulation components are complex and, most importantly, demand even more transfers from the physical space to the digital one: not only sizes (already a source of error~\cite{mahapatraBarriersEndUserDesigners2019,ramakersMeasurementPatternsUserOriented2023} -- fig. \ref{fig:results2}: 21, 26, 32) but also weight and weight distributions, for instance.
        In a sense, users were conducting tangible simulations within the object's context, by building brick assemblies, testing them, and altering them if needed.
        Lastly, we could observe a certain straightforwardness of the participants' use of \system, as they already knew its \textbf{language of expression}.
        While an introduction to \system or \tk was needed -- especially for novices -- none was required for the task of ``\textit{solving a mechanical challenge with plastic bricks}''.
        Users immediately started to ``block out'' geometry and test its function using relevant objects (i.e., plant, phone).
        Current design tools inherently require explicit learning: their interfaces, but also their paradigms (e.g., sculpting, programmatic, parametric, or solid modeling).
        For toy bricks, this is not the case -- expressing geometry through these bricks was natural to both novices and more experienced users in our study, in line with established benefits of tangible interaction~\cite{ishiiTangibleBitsPixels2008,horneckerGettingGripTangible2006}, and echoing insights from works employing malleable modeling materials~\cite{savageMakersMarksPhysical2015,jonesWhatYouSculpt2016} or other construction kits for 3D design~\cite{leenStrutModelingLowFidelityConstruction2017,andersonTangibleInteractionGraphical2000}.
        Similarly, \emph{testing} physical behavior in-situ with such low-fidelity models required little to no explanation -- especially when comparing this to more sophisticated ways to simulate how geometry behaves (e.g., the simulation workspace in Autodesk Fusion 360).
        With \system, we wanted users to be able to explore physical consequences early on while allowing them to \textbf{infer geometry} from 2D features.
        However, it is hard to holistically evaluate and explore these physical consequences with 2D outlines or their combinations: not all problems can be reliably solved in 2D or 2.5D.
        For instance, designing vases with a brick profile for the lathe reconstruction lets users omit the definition of the entire object.
        Detecting potential issues in balance or fit is then left to a purely digital preview, which has worked for some (e.g., \ref{fig:results2}: 10), but not all (e.g., \ref{fig:results2}: 14) designs made by the participants.
        Nevertheless, it is still possible and feasible for a subset of issues (i.e., ones that can be solved in 2D, such as the stands in figure \ref{fig:extrude}, \ref{fig:walkthrough-phonestand}, and \ref{fig:teaser}a).
        In the study, this has led to reasonably stable and fitting designs (e.g., fig. \ref{fig:results2}: 2, 4, 12).
        While the bricks provide a worst-case estimate of aspects like clearance, they may also induce a certain bias in the design process: constraints and approaches known from toy bricks may influence the users' designs, such as staggered layers of bricks (e.g., \ref{fig:results2}: 4 or fig. \ref{fig:results-prestudy}: 9).
        Users may also think that only a rigid brick model yields a rigid fabricated (e.g., printed) model, which is incorrect.
        This is not necessarily the case, as a loosely assembled brick model (e.g., fig. \ref{fig:walkthrough-bookend}a, or fig. \ref{fig:results2}: 2) still results in a solid model, while a solid brick assembly still likely indicates structural stability later on.
        A model that has completely disconnected parts, in turn, can be recognized as faulty early on.
        For novices, this was likely harder to do with \tk (resulting in designs like fig. \ref{fig:results2}: 27, 29).
        The aforementioned aspects likely have influenced our participants' designs (cf., section \ref{sec:prestudy} and \ref{sec:study}) and our own ones (cf., section \ref{sec:system} and \ref{sec:walkthroughs}).
        \system can support novices in designing everyday artifacts than interact with existing counterparts while ideally reducing the potential for errors related to measuring and physics (fig. \ref{fig:results2}: 7, 16), but is constrained to a specific grid, which either leads users to embed more clearance in their designs (fig. \ref{fig:results2}: 4, 12), or attempt to correct this digitally (fig. \ref{fig:results2}: 7, 10, 11).

%% file: sections/limitations.tex
 \section{Limitations}\label{sec:limitations}
    \paragraph{Technical Limitations}
    First and foremost, the scanning process does not always yield a perfect rendition of the tangible model, especially when users try to acquire side views of complete and complex (3D) objects, instead of profiles (2D).
    This limits the tool with respect to the geometries that are possible (e.g., complex cavities) and hinders insights gained with the tangible model (e.g., regarding stability), to generalize to the finished design, 
    
    \paragraph{Conceptual Limitations}
    \system also suffers from a certain \textbf{unidirectionality}, separating the tangible modeling phase from the remaining ones.
    The tangible building blocks are not synchronized with the digitally altered model. 
    To combat this limitation, this would require approaches as presented with ReForm~\cite{weichelReFormIntegratingPhysical2015} or continuously tracked bricks, where a tangible material is kept in sync with a digital representation.
    This would enable bidirectionality~\cite{kaimotoSketchedRealitySketching2022} between the tangible elements and the AR design environment of \system, and support more fluid modeling processes.
    In its current implementation, \system is also limited with respect to object \textbf{size} and can support objects that are above a brick's size.
    While it is possible to use small bricks to assemble large structures, their practical benefits diminish at larger scales.
    Large-scale modeling (cf.,~\cite{agrawalProtopiperPhysicallySketching2015,kovacsTrussFabFabricatingSturdy2017,kovacsTrusscillatorSystemFabricating2021}) would require a different medium, which ideally still supports tangible, in-situ design and testing.
    \paragraph{Evaluation}
        While our evaluation results indicate a certain usefulness of \system for novices, we must consider the study's small sample size, and uneven distribution of experts vs. novices.
        This prohibited us from conducting meaningful statistical analyses.
        Subsequently, we focused on errors, error types, and their distributions, outlining trends we observed in the data and in observations.
        We further considered designs that failed due to technical issues with \system to be non-functional, which limited the analysis of user error (presence or absence) in these cases, despite most brick assemblies that were created hinting at a valid design.
        Similarly, the labeling process may not have uncovered all possible issues of a design or identified the most evident, but potentially not the most mechanically crucial issue in a model.
        Technical and ergonomic issues (e.g., field of view) of the headset itself also affected the ratings (e.g., SUS) and success of users.
        
    \paragraph{Impact} 
        The design of low-effort design tools has to be considered under the umbrella of larger-scale effects~\cite{shneidermanSocialImpactStatements1996} of potential adoption~\cite{stemasovEphemeralFabricationExploring2022}, as its impact on the environment may likely be undesirable~\cite{goughDesignMouldGrow2023,wallScrappyUsingScrap2021}.
        Ideally, design tools for personal fabrication, operate either in reconfigurable media (e.g., bricks used by \system or comparable modules~\cite{suzukiDynablockDynamic3D2018}), create objects that can be deconstructed~\cite{wuUnfabricateDesigningSmart2020}, or operate purely digitally, to avoid the fabrication of frequent and permanent physical prototypes. 
        We consider \system to present an approach that embraces these principles while ideally reducing~\cite{muellerWirePrint3DPrinted2014,muellerDirectManipulationPersonal2018} or omitting~\cite{stemasovRoadUbiquitousPersonal2021} industrial notions of prototyping and iterating.

%% file: sections/futurework.tex
\section{Future Work}\label{sec:futurework}
    With \system, we have presented one approach to leverage low-fidelity tangible building materials for an in-situ design process.
    In the future, we intend to resolve the remaining technical frictions to be able to conduct a larger-scale user study. 
    This would also require the optimization and selection of specific interaction techniques (e.g., gestures vs. controller-based interactions) that facilitate the process outlined in \autoref{fig:process}.
    It is furthermore intriguing to consider whether we may use \emph{any} physical material to express designs and ideate in whichever material (or material combination) users want to or is available to them~\cite{piyaRealFusionInteractiveWorkflow2016,andersonTangibleInteractionGraphical2000,stemasovEnablingUbiquitousPersonal2021}.
    For future work, we are considering adding more input types to the system, such as clay~\cite{savageMakersMarksPhysical2015,jonesWhatYouSculpt2016}, 2D sketches~\cite{liSketch2CADSequentialCAD2020,saulSketchChairAllinoneChair2011}, 3D-printing pens~\cite{takahashi3DPen3D2019} or other construction kits such as K'Nex\footnote{\url{https://basicfun.com/knex/}, Accessed: 2.5.23} or Geomag\footnote{\url{https://www.geomagworld.com/en/}, Accessed: 29.4.23}, while retaining our notion of \emph{continuous} in-situ design.
    This would nudge \system further toward an approach that allows users a fully unrestricted choice of their medium of expression.
    Additional ways to infer geometry are conceivable for the reconstruction step, such as ``Teddy''~\cite{igarashiTeddySketchingInterface1999}.
    \system may also benefit from snapping: both to the initial block grid (e.g., when adding primitives later) or to the physical environment~\cite{nuernbergerSnapToRealityAligningAugmented2016}.
    Additional operations involving the physical environment in the design process (e.g., subtraction~\cite{stemasovMixMatchOmitting2020,zhuFusePrintDIY5D2016}) would also add expressivity to \system.
    Furthermore, the use of color-segmentation to extract profiles lays the groundwork for \emph{encoding} additional information in the tangible model: users may place specific colors of bricks to encode material properties or interactive elements to be automatically generated by \system (cf.,~\cite{savageMakersMarksPhysical2015,jonesWhatYouSculpt2016}), further decoupling the medium of expression from the medium of fabrication.

%% file: sections/conclusion.tex
\section{Conclusion}
    We presented \system, an in-situ, tangible modeling tool consisting of a mixed-reality application combined with tangible building blocks.
    It enables in-situ tangible design while affording early ideation, exploration, and iteration.
    In our study, we demonstrated that novice users made fewer errors related to balance and fit with \system, compared to \tk, and presented application walkthroughs highlighting a variety of approaches enabled through \system.
    
    Starting designs for fabrication in a tangible medium is meaningful and may serve as a \textit{``kickstart''} for subsequent, more digital design steps -- which should remain in-situ, instead of deferring error-prone transfers to later moments of the design process (cf.,~\cite{mahapatraBarriersEndUserDesigners2019,leenStrutModelingLowFidelityConstruction2017}).
    \system enables users to tangibly explore their designs early on while enabling them to refine them meaningfully without leaving the object's future physical context.
    All \emph{relevant} process steps (i.e., ideation, design, exploration, previewing) happen in-situ, at the location of the future artifact, instead of at spatially disconnected workstations.
    This aspect is further supported by the lower failure rate in balance and fit exhibited by the novice users (i.e., first-time 3D-modelers) in the study.
    We furthermore consider \system to be a step to decouple the medium of expression (for the user) and the medium of fabrication.
    By allowing users to express themselves in a highly approachable tool or material of their choice -- in our case, interlocking toy bricks -- users may benefit from an even lower entry barrier to 3D-design of artifacts to be fabricated, if the process is situated in the design's future space and context.

%% file: sections/acknowledgements.tex
\begin{acks}
    We thank the participants of our studies for their efforts and input. We further thank the reviewers of the manuscript for their suggestions.\\

    This research was partially funded by the Deutsche Forschungsgemeinschaft (DFG, German Research Foundation) through the project \textit{``Democratic and Sustainable Personal Design and Fabrication through In-situ Co-Design and Previsualization''} (project number: 525038300).
\end{acks}